\documentclass[12pt]{article}
\pdfoutput=1

\usepackage[utf8]{inputenc}
\usepackage[left=2.55cm, right=2.55cm, top=2.55cm, bottom=2.55cm]{geometry}
\usepackage{amsmath,amssymb,amsbsy}
\usepackage{slashed}
\usepackage{xcolor}
\definecolor{lightblue}{rgb}{0.93, 0.95, 1.0} 
\usepackage{graphicx}
\usepackage{cancel}
\usepackage{url}
\usepackage{cancel}
\usepackage{tcolorbox} 
\usepackage{cite}
\usepackage[colorlinks=true,allcolors=darkpurple,pdfborder={0 0 0},linktocpage=false]{hyperref}
\usepackage{tabularx,booktabs}
\usepackage{multicol}
\usepackage{units}
\usepackage{xspace}
\usepackage[labelfont=bf]{caption}
\usepackage[section]{placeins}
\usepackage{enumitem}
\usepackage{ulem}
\usepackage{subcaption}
\usepackage{multirow}
\usepackage{comment}
\usepackage{verbatim}
\usepackage{ulem}
\usepackage{comment}
\useunder{\uline}{\ul}{}
\usepackage{array} 
\newcolumntype{C}[1]{>{\centering\arraybackslash}p{#1}}
\renewcommand{\arraystretch}{1.5}
\usepackage[compat=1.1.0]{tikz-feynman}
\definecolor{darkred}{rgb}{0.6,0,0}
\definecolor{darkpurple}{rgb}{0.5,0,0.5}

\def\abs[#1]{\lvert #1\rvert}
\def\L{\mathcal{L}}
\def\M{\mathcal{M}}
\def\O{\mathcal{O}}
\def\D{\mathcal{D}}
\def\hc{\text{h.c.}}
\def\Y{\mathcal{Y}}
\def\BR{\text{BR}}
\def\IM{\, \text{Im}}

\def\z2{$\mathbb{Z}_2$}
\def\z3{$\mathbb{Z}_3$}

\def\id{\mathbb{I}}

\def\U1L{$\mathrm{U(1)}_L$}
\newcommand{\dneff}{\Delta N_{\text{eff}}}
\newcommand{\SM}{\text{SM}}

\newcommand{\abst}[1]{\left| #1 \right|}

\definecolor{avblue}{rgb}{0.0, 0.0, 0.8}
\definecolor{asparagus}{rgb}{0.53, 0.66, 0.42}
\definecolor{aqua}{rgb}{0.4, 0.6, 0.7}
\definecolor{thblue}{RGB}{174, 216, 235}

\newcommand{\scch}[1]{{\color{asparagus} #1}}

\newcommand {\ignore}[1]{}

\newcommand{\AddrIFIC}{%
  Instituto de F\'{i}sica Corpuscular, CSIC-Universitat de Val\`{e}ncia, 46980 Paterna, Spain}

\newcommand{\AddrHeidelberg}{%
  Max-Planck-Institut f\"{u}r Kernphysik, Saupfercheckweg 1, 69117 Heidelberg, Germany}

\newcommand{\AddrFISTEO}{%
  Departament de F\'{\i}sica Te\`{o}rica, Universitat de Val\`{e}ncia, 46100 Burjassot, Spain}

\begin{document}


\begin{center}
\vspace*{15mm}

\vspace{1cm}
{\Large \bf 
Flavour and cosmological probes of Diracon models
} \\
\vspace{1cm}

{\bf Salvador Centelles Chuli\'a$^{\text{a,b}}$, \bf Tim Herbermann$^{\text{a}}$,\\ Antonio Herrero-Brocal$^{\text{b}}$, Avelino Vicente$^{\text{b,c}}$}

\vspace*{.5cm}
 $^{(\text{a})}$ \AddrHeidelberg \\\vspace*{.2cm} 
 $^{(\text{b})}$ \AddrIFIC \\\vspace*{.2cm} 
 $^{(\text{c})}$ \AddrFISTEO

\vspace*{.3cm}
\href{mailto:salcen@ific.uv.es}{salcen@ific.uv.es},
\href{mailto:tim.herbermann@mpi-hd.mpg.de}{tim.herbermann@mpi-hd.mpg.de},\\
\href{mailto:antonio.herrero@ific.uv.es}{antonio.herrero@ific.uv.es},
\href{mailto:avelino.vicente@ific.uv.es}{avelino.vicente@ific.uv.es}
\end{center}

\vspace*{0mm}
\begin{abstract}\noindent\normalsize

We present and analyze two minimal extensions of the Standard Model featuring a spontaneously broken global, chiral, and anomaly-free $U(1)_D$ symmetry. This breaking generates naturally small Dirac neutrino masses via a seesaw mechanism and yields a physical massless Goldstone boson, the Diracon. Although both models share the same particle content and scalar potential, their distinct symmetry breaking pattern leads to remarkably different phenomenological and cosmological signatures. In the first model, the Diracon couples weakly to charged leptons but right-handed neutrinos can be efficiently produced  in the early Universe, resulting in stringent constraints from the effective number of relativistic species, $\Delta N_{\text{eff}}$. Conversely, in the second one, right-handed neutrino production is suppressed, and flavour-violating processes such as $\mu \to e \D$ provide the most promising probes. These simple but elegant models showcase the complementarity between cosmological observations and low-energy flavour experiments in the search for physics beyond the Standard Model. 
\end{abstract}

\section{Introduction}
\label{sec:intro}
The absence of neutrino masses in the Standard Model (SM) is one of its most well-known shortcomings, especially in light of the discovery of neutrino oscillations~\cite{Kajita:2016cak, McDonald:2016ixn}. The large gap between the Electroweak (EW) scale and the neutrino mass scale provides a compelling window into potential new physics related to the mechanism of mass generation. Even more promising is the fact that neutrinos are electrically neutral under the SM gauge group, which allows for the inclusion of a Majorana mass term if right-handed (RH) neutrinos are introduced to account for their masses. Based on this reasoning, the conventional Type-I seesaw mechanism~\cite{Minkowski:1977sc,Yanagida:1979as, Mohapatra:1979ia,Gell-Mann:1979vob, Schechter:1980gr} naturally appears as an elegant explanation for neutrino masses by introducing a very large Majorana mass term for the RH neutrinos.

However, new symmetries under which neutrinos are charged could emerge as fundamental symmetries, protecting their Dirac nature. One such example is an accidental global symmetry already present in the SM, often referred to in the literature as lepton number. In this work, we instead denote it as $U(1)_D$. This reinterpretation allows for a generalized and elegant extension of the seesaw mechanism to include Dirac neutrino scenarios~\cite{Ma:2014qra, Ma:2015mjd,Ma:2015raa,Ma:2016mwh, Chulia:2016ngi}. Although the Dirac neutrino scenario has historically received limited attention, it has garnered increased interest in recent years. For instance, see~\cite{CentellesChulia:2016fxr,CentellesChulia:2017koy,Bonilla:2018ynb,CentellesChulia:2018bkz,CentellesChulia:2019xky,Peinado:2019mrn,Wang:2017mcy,Borah:2017leo,Jana:2019mgj,Jana:2019mez,Calle:2019mxn,Nanda:2019nqy,Ma:2019byo,Ma:2019iwj,Correia:2019vbn,Saad:2019bqf,Ma:2019yfo,CentellesChulia:2020dfh,CentellesChulia:2020bnf,Guo:2020qin,delaVega:2020jcp,Borgohain:2020csn,Leite:2020wjl,Chulia:2021jgv,Bernal:2021ppq,Mishra:2021ilq,Biswas:2021kio,Mahanta:2021plx,Hazarika:2022tlc,CentellesChulia:2022vpz,Berbig:2022nre,Maharathy:2022gki,Chowdhury:2022jde,Biswas:2022vkq,Li:2022chc,Berbig:2022hsm,Mahapatra:2023oyh,Borah:2024gql,CentellesChulia:2024iom,Dey:2024ctx,Singh:2024imk,Kumar:2025cte,Batra:2025gzy} for a non-exhaustive selection of relevant examples.

The spontaneous symmetry breaking (SSB) pattern of $U(1)_D$ has important phenomenological consequences. For example, the remnant subgroup determines the Dirac/Majorana nature of neutrinos~\cite{Hirsch:2017col} or can stabilize a dark matter (DM) candidate~\cite{CentellesChulia:2016rms,Bonilla:2018ynb, CentellesChulia:2018gwr,CentellesChulia:2019gic}. More importantly, it implies the existence of a Nambu--Goldstone boson in the spectrum and determines its interactions: the Majoron ($J$)~\cite{Chikashige:1980qk,Chikashige:1980ui,Schechter:1981cv,Gelmini:1980re,Aulakh:1982yn, Pilaftsis:1993af, CentellesChulia:2024uzv} in the Majorana case, or the Diracon ($\D$)~\cite{Bonilla:2016zef,Bonilla:2016diq,CentellesChulia:2018gwr,Berbig:2023uzs} in the Dirac case. In either scenario, this massless boson has important phenomenological consequences and can be probed through various experiments, from rare low-energy processes~\cite{Escribano:2020wua} to invisible Higgs decays~\cite{Joshipura:1992hp} at high-energy colliders, as well as through its imprint on cosmological observables~\cite{Weinberg:2013kea,AristizabalSierra:2014uzi,Bonilla:2019ipe,Escudero:2019gvw,DeRomeri:2022cem}.

Focusing on the Dirac neutrino possibility, if the light RH neutrinos are abundantly produced in the early Universe, their energy density contributes to its total expansion rate~\cite{Adshead:2020ekg, Luo:2020fdt, Luo:2020sho}, and this is tightly constrained by observations of the cosmic microwave background (CMB) and big bang nucleosynthesis (BBN). Models that protect the Dirac nature of neutrinos by means of a gauge symmetry have been thoroughly studied and face tight and rather generic constraints, see e.g. Refs.~\cite{Heeck:2014zfa,Abazajian:2019oqj,Adshead:2022ovo,Herbermann:2025uqz,Borah:2024twm,Ghosh:2024cxi}. On the other hand, ensuring the Dirac nature through a global symmetry leads to model-dependent cosmological constraints, see e.g. \cite{Borah:2022enh,Biswas:2022fga,Berbig:2022pye,Biswas:2024gtr,Borah:2023dhk,Borboruah:2024lli,Borah:2025fkd}. Here we focus on the Dirac Type-I seesaw scenario as a natural extension of the framework developed in~\cite{CentellesChulia:2024uzv} for Majorana seesaw models and study the flavour and cosmological consequences of the SSB of the $U(1)_D$ symmetry.

The paper is organized as follows. In Sections~\ref{sec:Type-I} and \ref{sec:models} we define the Dirac Type-I seesaw family, discuss a minimal scheme for spontaneous symmetry breaking and introduce two viable realizations. Sections~\ref{sec:Flavour} and \ref{sec:Cosmo} discuss the general phenomenological implications of this class of models in flavour observables and cosmological probes, respectively. In Section~\ref{sec:pheno}, we explore the specific phenomenology of each realization, showing that while the canonical Dirac Type-I seesaw is constrained by cosmological observations, the alternative model remains cosmologically untested and offers promising signatures in Diracon-mediated lepton flavour violation (LFV) processes. Finally, we present our conclusions in Section~\ref{sec:summary}. Additional details are provided in Appendices~\ref{app:coupling}, \ref{app:scalar} and \ref{app:Yukawa}.

\section{The Dirac Type-I seesaw family}
\label{sec:Type-I}
We start by defining the Dirac Type-I seesaw family. For a similar discussion in the Majorana case we refer to~\cite{CentellesChulia:2024uzv}. A model belongs to the Dirac Type-I seesaw family if it satisfies three conditions: 
\begin{itemize}
\item Its neutral fermion mass Lagrangian can be written as
\begin{equation}
	\L_{\textup{mass}} = \begin{pmatrix}
	 \bar{\nu}_L & \bar{N}_L 
\end{pmatrix}	
\begin{pmatrix}
	0 & M_1\\
	M_2 & M_N
\end{pmatrix} 
\begin{pmatrix}
	\nu_R \\
	N_R
	\end{pmatrix} + \hc \equiv 
\begin{pmatrix}
	 \bar{\nu}_L & \bar{N}_L 
\end{pmatrix}	
	\M \begin{pmatrix}
	\nu_R \\
	N_R
	\end{pmatrix} + \hc\, , \label{massmatrix}
\end{equation}
where $\nu_L$ are the usual 3 SM neutrinos while $\nu_R$ and $N_{L,R}$ are new beyond the SM (BSM) neutral fermions. We consider $n_{\nu_R} \leq 3$ generations of $\nu_R$ and $n_N$ generations of $N_{L,R}$.~\footnote{Scenarios with $n_{\nu_R} > 3$ include unpaired right-handed neutrinos which may remain massless after symmetry breaking. While these models maybe be phenomenologically viable and interesting, we will not include them in our analysis.}
\item The hierarchies $\left( M_1 \, M_N^{-1} \right)_{ij} \ll 1$ and $\left( M_2 \, M_N^{-1} \right)_{ij} \ll 1 $ $\forall i, j$ are satisfied. We call them the seesaw conditions. They allow us to identify the seesaw expansion parameters, $\epsilon_1 = \O (M_1 M_N^{-1})$ and $\epsilon_2 = \O (M_2 M_N^{-1})$.
\item All possible Majorana neutrino mass terms are forbidden by symmetry.
\end{itemize}
These three conditions define a general setup that leads to a naturally small Dirac mass term for three light neutrinos, $M_\nu \ll \Lambda_{\textup{EW}}$, with $\Lambda_{\textup{EW}}$ denoting the EW scale. One can derive a general formula for this Dirac mass by rotating from the flavour to the mass basis. This rotation is performed by two unitary matrices: $U_L$, which acts on the left-handed fermions, and $U_R$, which acts on the right-handed ones. They are defined by 
\begin{equation}
U_L^\dagger \, \M \, U_R = \hat{\M}= \text{diag}\left(m_1,m_2, \dots, m_{3+n_N}\right) \, ,
\end{equation}
where $m_1, m_2, m_3 \ll \Lambda_{EW} \ll m_4... m_{3+n_N}$ if the seesaw conditions are satisfied. $U_{L,R}$ are two unitary matrices of dimensions $(3+n_N) \times (3+n_N)$ and $(n_{\nu_R}+n_N) \times (n_{\nu_R}+n_N)$, respectively, which can be written as
\begin{equation} \label{eq:Ufactor}
    U_{L,R}= \begin{pmatrix}
    \sqrt{\id_{3, n} - P_{L,R} P_{L,R}^\dagger} & P_{L,R} \\
    -P_{L,R}^\dagger & \sqrt{\id_{n} - P_{L,R}^\dagger P_{L,R}}
    \end{pmatrix}\, \begin{pmatrix}
    (U_\ell)_{L,R} & 0 \\
    0 & (U_h)_{L,R}
    \end{pmatrix} \, .
\end{equation}
The left-handed light/heavy block rotations, $(U_\ell)_L \equiv U_\ell$ and $(U_h)_L$ have dimensions $3\times 3$ and $n_N \times n_N$, respectively, while the auxiliary (rectangular in general) matrix $P_L$ has dimensions $3 \times n_N$. In the seesaw expansion the first term block-diagonalizes the neutrino mass matrix into the light and heavy blocks, which are later diagonalized by $U_\ell$ and $(U_h)_L$, respectively. Thus, at first order in the seesaw expansion we identify $U_\ell$ with the usual lepton mixing matrix relevant for neutrino oscillations. Similarly, the right-handed unitary matrices $(U_\ell)_R$ and $(U_h)_R$ have dimensions $n_{\nu_R}\times n_{\nu_R}$ and $n_N \times n_N$ respectively, while $P_R$ is $n_{\nu_R}\times n_N$. We denote the identity matrix of rank $m$ as $\id_m$.
The block-diagonalization can be performed in a similar way to the Majorana case. We can find a general solution by expressing the matrices $P_{L,R}$ as power series of the seesaw expansion parameters. By doing so, we find at leading order
\begin{align}
P_L = M_1 \,M_N^{-1} \, , && P_R^\dagger = M_N^{-1} \, M_2 \, .
\end{align}
This leads to a light neutrino mass matrix given by
\begin{equation} \label{eq:Mnu}
M_\nu = - M_1 \, M_N^{-1} \, M_2 \, ,
\end{equation}
and diagonalized by $U_\ell$ and $U_{\ell R}$ as $\widehat M_\nu = U_\ell^\dagger M_\nu U_{\ell R}$. This allows us to easily parametrize the matrix $M_1$ in terms of the physical parameters $M_N$, $U_\ell$ and the neutrino masses $\widehat M_\nu$ as well as the model parameters $M_2$ and $U_{\ell R}$,
\begin{equation} \label{eq:parametrization}
    M_1 = - U_\ell^\dagger \, \widehat M_\nu \, U_{\ell R} \, M_2^{-1} \, M_N \, .
\end{equation}
Let us make some comments on the expression for the light neutrino mass matrix given in Eq.~\eqref{eq:Mnu}. First, this equation generalizes the one found for the Majorana scenario~\cite{CentellesChulia:2024uzv}, where $\nu_R = \nu_L^c$, $N_R = N_L^c$ and, therefore, $M_2 = M_1^T$. Just as in the Majorana case, Eq.~\eqref{eq:Mnu} holds regardless of the structure of $M_1$, $M_N$ and $M_2$, i.e. the new BSM neutral fermions, $N_{R,L}$ could belong to multiplets with different charges under the symmetries of a UV-complete model. This implies that Eq.~\eqref{eq:Mnu} not only reproduces the result in models in which $N$ correspond to $n_N$ generations of a single type of BSM neutral fermion (as in~\cite{Ma:2015mjd,Ma:2014qra,Chulia:2016ngi,CentellesChulia:2018gwr}) but also in models that feature different types of BSM fermions, such as those appearing in Dirac double or triple inverse seesaw models~\cite{CentellesChulia:2020dfh}.
On the other hand, it is important to note that one can express the result as $M_\nu = - \epsilon_1 M_2$ where $\epsilon_1$ represents the mixing between the SM neutrinos and the new heavy neutral fermions. In models in which all the BSM neutral fermions introduced are singlets under the SM gauge group, the effects of new physics (NP) arise solely from the mixing, i.e. $\epsilon_1$. In canonical \textbf{high-scale seesaw scenarios}, Majorana or Dirac, the smallness of neutrino masses relies on a suppressed mixing. However, this also leads to very suppressed NP effects. Alternatively, one may take advantage of the $M_2$ matrix, which also enters the neutrino mass equation. Naively,
\begin{equation}
\epsilon_1 \sim \frac{M_\nu}{M_2} \, , \label{eq:lowscalegeneral}
\end{equation}
and the mixing depends on the scale of $M_2$. For instance, a sizable mixing is possible when $M_2 \ll M_1 \ll M_N$. Scenarios of this class, which we refer to as \textbf{low-scale seesaw scenarios}, succeed in explaining neutrino masses while preserving the possibility of observable NP signals.

Finally, we note that in order to realize a model of the Dirac Type-I seesaw family in a consistent way, the Dirac nature of neutrinos must be protected by a symmetry which forbids all Majorana mass terms. Similarly, the tree-level mixing between $\nu_L$ and $\nu_R$ must also be forbidden by a symmetry argument. While there exist infinitely many possibilities for new BSM symmetries and fields to achieve these features, in this work we will focus on two minimal models, which we proceed to present in the next Section.
\section{Two minimal Dirac type-I seesaw models}
\label{sec:models}

We now introduce two models that feature a global, chiral and anomaly-free $U(1)_D$ symmetry. This Abelian symmetry is multipurpose and well motivated, having the following attractive features:

\begin{enumerate}[label=(\roman*)]
    \item Forbids all Majorana mass terms, thus protecting the Dirac nature of neutrinos~\cite{Hirsch:2017col}.
    \item The chiral charges between $\nu_L$ and $\nu_R$ forbid their tree-level mass term, which has to come at a higher-order operator, thus naturally explaining the smallness of neutrino mass~\cite{CentellesChulia:2018gwr}.
    \item It is anomaly-free if identified with $U(1)_{B-L}$, thus one could immediately promote $U(1)_D$ to a gauge symmetry~\cite{Montero:2007cd, Ma:2014qra}. Here we focus on the global case, leading to a massless Goldstone.
    \item It is well known that such a $U(1)_D$ can also stabilize a DM candidate~\cite{Bonilla:2018ynb}. We do not explicitly analyze this possibility here.
\end{enumerate}

Both models also add the same particles to the SM: three new types of neutral fermions, $N_L$, $N_R$ and $\nu_R$, and a complex scalar $\sigma$ which will spontaneously break $U(1)_D \to \mathbb{Z}_3$. The difference between the two models comes at the $U(1)_D$ symmetry charges, which are given in Table~\ref{tab:fields}. In \textbf{Model I}, which will also be referred to as \textbf{Canonical model}, three $\nu_R$ singlets are introduced, charged under $U(1)_D$ as $(-4,-4,\,5)$. In addition, two generations of $N_{L,R}$ are considered. In \textbf{Model II},  three generations of $\nu_R$ and $N_{L,R}$ are introduced. This model will also be called \textbf{Enhanced Diracon model}, for reasons to be understood below. In this case the $U(1)_D$ charges of the three $\nu_R$ singlets are universal, while the charges of $N_L$ and $N_R$ are different. This slight variation of the symmetry assignment in each model leads to very distinct phenomenology, as will be shown in what follows. Finally, the quark sector of both models is assumed to be as in the SM.

\begin{table}[tb!]
\begin{center}
\renewcommand{\arraystretch}{1.35} %
\setlength{\tabcolsep}{5pt} 
\begin{tabular}{| c | c | c | c |}
\multicolumn{2}{c}{} & \multicolumn{1}{c}{\textbf{Canonical}}   & \multicolumn{1}{c}{\textbf{Enhanced Diracon}}\\
\hline
Fields & $SU(2)_L \otimes U(1)_Y$ & $U(1)_D$  & $U(1)_D$ \\ 
\hline
$H$ & $ (\textbf{2}, \frac{1}{2}) $ & 0 & 0 \\ 
$\sigma$ & $(\textbf{1}, 0)$ & $3$ & $3$ \\
\hline
$L$ & $(\textbf{2}, -\frac{1}{2})$ & -1  & -1\\
$\nu_R$ & $ (\textbf{1}, 0) $ & $(-4,-4,\,5)$ & $2$ \\
$N_L$ & $ (\textbf{1}, 0) $ & -1 &$2$ \\
$N_R$ & $ (\textbf{1}, 0) $ & -1 & -1 \\
\hline
\end{tabular}
\end{center}
\caption{Global $U(1)_D$ and gauge electroweak charges of the particles in the two minimal Dirac type-I seesaw models. The $D$ charge of $N_R$ is fixed by the $M_1$ term in Eq.~\eqref{massmatrix}, while the $D$ charge of $\sigma$ is model-dependent and sequentially fixes the charges of $N_L$ and $\nu_R$. Here we choose the simplest case, with a charge of $3$. 
\label{tab:fields}}
\end{table}


\subsection{Scalar sector}
\label{subsec:Scalar}
Within this minimal scheme the scalar sector will be formally identical in both models. The most general potential under the symmetry group $SU(3)_C \times SU(2)_L \times U(1)_Y \times U(1)_D$ is given by
\begin{align}
 \mathcal V (H, \sigma) = -\mu^2 (H^\dagger H)- \mu^2_\sigma (\sigma^\dagger \sigma)+\lambda (H^\dagger H)(H^\dagger H) + \lambda_\sigma (\sigma^\dagger \sigma)(\sigma^\dagger \sigma) +\lambda_{H \sigma} (H^\dagger H)(\sigma^\dagger \sigma) \, .
\end{align}
Assuming CP conservation in the scalar sector, the SM Higgs doublet $H$ and the singlet $\sigma$ can be decomposed as
\begin{align}
H = \begin{pmatrix}
G_W^+ \\
\frac{1}{\sqrt{2}} \left( v + S_H + i \, G_Z \right)
\end{pmatrix} \, , && \sigma = \frac{1}{\sqrt{2}} \left( v_\sigma + S_\sigma + i \, \D \right) \, .
\end{align}
Here $\langle H^0 \rangle = v/\sqrt{2}$ and $\langle \sigma \rangle = v_\sigma/\sqrt{2}$ are the vacuum expectation values (VEVs) of the neutral component of $H$ and $\sigma$, respectively. The former is responsible for the breaking of the electroweak symmetry, just as in the SM, while the latter breaks the $U(1)_D$ global symmetry.

Let us now discuss the resulting scalar spectrum after symmetry breaking. In the neutral CP-even sector, we find two massive states. In the basis $S = \left( S_H, S_\sigma \right)$, the scalar mass matrix is given by
\begin{equation}
    \mathcal M_S^2 = \begin{pmatrix}
        2 v^2 \,\lambda & v_\sigma v \, \lambda_{H\sigma} \\
        v_\sigma v \,\lambda_{H\sigma} & 2 v_\sigma^2 \,\lambda_\sigma
    \end{pmatrix} \, ,
\end{equation}
which can be diagonalized as
\begin{align}
U^\dagger \, \mathcal M_S^2 \, U = \begin{pmatrix}
       m_h^2 & 0 \\
       0 & m^2_{\mathcal{S}}
       \end{pmatrix} \,, &&
   \text{with} \,\,\, U=  \begin{pmatrix} 
     \cos \alpha & \sin \alpha \\
       -\sin\alpha & \cos \alpha 
   \end{pmatrix} \, .
\end{align}
It proves convenient to rewrite the potential parameters in terms of $m_h$, $m_\mathcal{S}$ and $\alpha$. One obtains
\begin{align}
    2 v^2 \lambda = \cos^2 \alpha \, m_h^2 + \sin^2 \alpha \, m_\mathcal{S}^2 \, ,\\
    2 v_\sigma^2 \lambda_\sigma = \sin^2 \alpha \, m_h^2 + \cos^2 \alpha \, m_\mathcal{S}^2 \, ,\\
    v_\sigma v \lambda_{H \sigma} = \cos \alpha \, \sin \alpha \, (m_\mathcal{S}^2 - m_h^2) \, .
\end{align}
In the small mixing limit, $\alpha \ll 1$, we identify $h \approx S_H$ as a SM-like Higgs of $m_h \approx 125$ GeV. The spectrum contains an additional massive scalar, $\mathcal{S}$. The mass of this new \textit{radial scalar} is given by $m_\mathcal{S} \sim v_\sigma$. On the other hand, in the charged and neutral CP-odd sector we find, as expected, the three unphysical Goldstone bosons, $G^+_W$ and $G_Z$, which give rise to the longitudinal component of the SM gauge bosons $W^+$ and $Z$, and one physical pure singlet Goldstone boson, $\D$, the Diracon~\cite{Bonilla:2016zef}, due to the spontaneous breaking of the global $U(1)_D$ symmetry, which constitutes one of the main features of these two models. In general, the Diracon has significant phenomenological implications, to be probed by different experiments. These include searches for rare low-energy processes like $\mu \to e \D$,  high-energy collider signatures such as invisible Higgs decays, and cosmological probes. In both models, the Diracon contribution to the Higgs invisible decay is given by~\cite{Joshipura:1992ua,Joshipura:1992hp}
\begin{equation}
    \Gamma(h \to \mathcal{D} \mathcal{D}) = \frac{m_h^3 }{32 \pi} \frac{\sin^2 \alpha}{v^2_\sigma}  \, .
\end{equation}
If kinematically allowed, the Higgs boson can also decay to a pair of $\mathcal{S}$ scalars, with a decay width approximately given by
\begin{equation}
    \Gamma(h \to \mathcal{S} \mathcal{S}) \approx \frac{\left(m_h^2+2 m_\mathcal{S}^2\right)^2}{32 \pi m_h} \frac{\sin^2 \alpha}{v^2_\sigma}  \sqrt{1-4\frac{m_\mathcal{S}^2}{m_h^2}}\, .
\end{equation}
The decay of the $\mathcal{S}$ scalars is invisible, since it only couples to neutrinos (see Sec.~\ref{subsec:YukMnu} below). Thus, we obtain the limit
\begin{equation}
  \Gamma(h \to \mathcal{D} \mathcal{D})+ \Gamma(h \to \mathcal{S} \mathcal{S})  < \Gamma^\text{inv} \, ,
\end{equation}
where $\Gamma^\text{inv}$ is the 95\% CL limit on the invisible Higgs decay, currently given by $\Gamma^\text{inv} < 0.39$ MeV~\cite{ParticleDataGroup:2024cfk}. The scattering amplitudes relevant for cosmology are approximately given by
\begin{align}
\mathcal{A}(h h \to \mathcal{D}\mathcal{D}) \approx& \frac{\sqrt{\sqrt{2}G_F}}{4} \, m_h^2 \, \frac{\sin\alpha}{v_\sigma} \left[  \left(2+ \frac{m_\mathcal{S}^2}{m_h^2}\right) \frac{m_\mathcal{S}^2}{s-m_\mathcal{S}^2 + i m_\mathcal{S} \Gamma_\mathcal{S}} -1+\frac{m_\mathcal{S}^2}{m_h^2}-\frac{m_h^2}{s-m_h^2 + i m_h \Gamma_h}\right] \, , \nonumber \\
\mathcal{A}(h h \to \mathcal{S} \mathcal{S}) \approx& -\frac{\sqrt{\sqrt{2}G_F}}{4} \, m_h^2 \, \frac{\sin\alpha}{v_\sigma} \left[\frac{m_h^2+2 m_\mathcal{S}^2}{s-m_h^2 + i m_h \Gamma_h}+1-\frac{m_\mathcal{S}^2}{m_h^2}\right] \, ,
\end{align}
while other relevant analytical results are relegated to Appendix~\ref{app:scalar}.

\subsection{Yukawa interactions and neutrino masses}
\label{subsec:YukMnu}

We now discuss the Yukawa interactions and the generation of neutrino masses in both models.

\subsubsection*{Canonical model}

The Yukawa Lagrangian of this model is given by
\begin{equation}
    \L_Y^{I} =  Y\, \bar{L} \, \tilde{H} N_R + Y' \bar{N}_L \sigma \, \nu_R^{(1,2)} + M_N \bar{N}_L \, N_R + \hc \, .
\end{equation}
Due to the $U(1)_D$ charges, the model features only $2$ coupled right-handed neutrinos, which we denote as $\nu_R^{(1,2)}$. For this reason, the matrices $Y'$ and $M_N$ are $2\times 2$ while $Y$ is $3\times 2$ and therefore one of the active neutrinos is massless. The right-handed neutrino with charge $5$ under $U(1)_D$ is also massless, but does not have any interaction.~\footnote{One can build a completely equivalent version of the canonical model by removing the third right-handed neutrino, with charge $5$ under $U(1)_D$. However, we have decided to keep it to enforce the cancellation of the $U(1)_D$ anomalies. While this is not necessary in our case, since we have taken $U(1)_D$ to be a global symmetry, it would be relevant for gauged variants of this model or extensions with three massive neutrinos.} The neutral fermion mass matrix is given by
\begin{equation}
    \mathcal{M}_n = \begin{pmatrix}
        0 & \frac{v}{\sqrt{2}} Y \\
        \frac{v_\sigma}{\sqrt{2}} Y' & M_N\\
    \end{pmatrix} \, .
\end{equation}
Assuming the seesaw conditions, which in this case read $Y v \, , Y' v_\sigma \ll M_N$, and following Eqs.~\eqref{eq:Mnu} and \eqref{eq:parametrization}, the $3\times 2$ light neutrino mass matrix can be computed as
\begin{equation}
M_\nu = \frac{v \, v_\sigma}{2} \, Y \, M_N^{-1} \, Y'  \, .
\label{eq:Mnucanonical}\end{equation}
This allows us to write the $3\times 2$ Yukawa marix $Y$ as
\begin{align} \label{eq:CasasIbarraCanonical}
   Y =  \frac{2}{v \, v_\sigma} \, \, U_\ell \, \widehat M_\nu \, U_{\ell R}^\dagger \, (Y')^{-1} \,M_N  \, .
\end{align}
We note that the seesaw conditions naturally lead to the existence of two light Dirac neutrinos (in addition to the massless one). In fact, neutrino masses are suppressed by the largest scale in the model, $M_N$. $\widehat M_\nu$ has a row of zeroes and a diagonal $2\times 2$ block. For example, for normal ordering and taking the best-fit values for the $\Delta m^2_{ij}$ oscillation parameters~\cite{deSalas:2020pgw}, $\widehat M_\nu$ is given by 
\begin{equation}
    \widehat M_\nu \approx \begin{pmatrix}
        0 & 0 \\
        0.0086 & 0 \\
        0 & 0.0504
    \end{pmatrix} \, \text{eV} \, .
\end{equation}
Without loss of generality we can work in the basis where the charged lepton mass matrix $M_\ell$ and $M_N$ are diagonal, real and positive.
\ignore{We can now make the simplifying assumption that the heavy neutral fermions are degenerate in mass with $(M_N)_{11} = (M_N)_{22} = M$. This choice gives us the option to work in the basis where $Y'$ is diagonal without loss of generality. In turn, if we also assume the singular values of $Y'$ to be degenerate, i.e. $Y'_{11} = Y'_{22} = y'$, one finds 
\begin{equation}
    Y = \frac{2  \, M}{y' \, v \, v_\sigma} \, U_\ell \, M_\nu \, U_{\ell R}^\dagger \, .
    \label{eq:canonical_Yuk}
\end{equation}
While the numerical scans performed in Sec.~\ref{sec:pheno} will be done for general $M_N$ and $Y'$, these simplifications will allow for very simple analytical results in what follows.}

Following the discussion around Eq.~\eqref{eq:lowscalegeneral}, the low-scale regime of the model corresponds to \( v_\sigma \ll v \). In this regime, the mixing between the active neutrinos and the heavy neutral fermions can be sizeable, giving rise to rich phenomenological signatures, as we will discuss below. Conversely, if \( v_\sigma \) is comparable to or larger than \( v \), the model enters the high-scale regime and the associated phenomenology becomes suppressed.

\subsubsection*{Enhanced Diracon model}

In this model, the charges of the fields under $U(1)_D$ allow us to write the Yukawa Lagrangian
\begin{equation}
    \L_Y^{II} =  Y\, \bar{L} \, \tilde{H} N_R + M_2 \, \bar{N}_L \, \nu_R + Y' \bar{N}_L \, \sigma \, N_R + \hc \, ,
\end{equation}
and, thus, the neutral fermion mass matrix is given by
\begin{equation}
    \mathcal{M}_n = \begin{pmatrix}
        0 & \frac{v}{\sqrt{2}} Y \\
        M_2 & \frac{v_\sigma}{\sqrt{2}}Y' \\
    \end{pmatrix} \, .
\end{equation}
We may now impose the seesaw conditions, $Y v \, , M_2 \ll Y' v_\sigma$, and again use Eqs.~\eqref{eq:Mnu} and \eqref{eq:parametrization} to find
\begin{equation}\label{eq:Casasenhanced}
M_\nu = -\frac{v}{v_\sigma} Y \, (Y')^{-1} \, M_2 \quad \to \quad Y = - \frac{v_\sigma}{v} \, U_\ell \, \widehat M_\nu \, U_{\ell R}^\dagger \, M_2^{-1} \, Y' \, .
\end{equation}
Again, the seesaw conditions lead to naturally small neutrino masses, even if the Yukawa couplings $Y$ and $Y'$ are sizable. In this model, the low-scale regime corresponds to the limit $M_2 \ll Y v$, leading to observable BSM phenomena. Deviating from this regime would lead instead to negligible deviations from the SM phenomenology, which we call the high-scale regime.

Compared to the canonical model, here $\sigma$ couples to $\bar N_L \, N_R$, while the bilinear $\bar N_L \, M_2 \, \nu_R$ is a bare mass term. The seesaw condition $M_2 \ll M_N$ implies that the effective mass term corresponding to the $\bar N_L \, \nu_R$ bilinear will be much smaller than that corresponding to $\bar N_L \, N_R$. Consequently, while all NP effects arising from mixing with the SM remains unaffected, the interaction introduced by $\sigma$, and in particular the physics associated with the Diracon, changes entirely, as we will show in Section~\ref{sec:pheno}.

\subsection{A comment on lepton number}
\label{subsec:lepton}

We conclude this Section with a comment on lepton number. Since the $U(1)_D$ charges of $L$ and $H$ are $-1$ and $0$, respectively, Ref.~\cite{Bonilla:2018ynb} identified this symmetry with $B-L$. This is a valid choice. However, we have preferred to avoid a potential confusion. The models discussed here lead to Dirac neutrino masses and, as a consequence of this, both lepton and baryon numbers are conserved at the perturbative level. One can easily check that this is true in both models. In the case of baryon number this is obvious, since both models have exactly the same quark sector as the SM. In the case of lepton number, one can find a charge assignment that preserves it, even after symmetry breaking. If we denote the lepton number symmetry as $U(1)_\ell$, these charges are $\ell_L = \ell_{\nu_R} = \ell_{N_L} = \ell_{N_R} = 1$ and $\ell_H = \ell_\sigma = 0$.~\footnote{Actually, the lepton number of the third right-handed neutrino in the canonical model is undetermined, since this fermion does not couple to any other field.} We note, however, that our $U(1)_D$ symmetry is not equivalent to this $U(1)_\ell$ lepton number symmetry. For instance, if we replaced $U(1)_D$ by $U(1)_\ell$, $\sigma$ would become a real singlet, since all its charges would be zero. In this case, a cubic $H^\dagger H \sigma$ term would be allowed in the scalar potential, with possible phenomenological consequences.
\section{Flavour observables}
\label{sec:Flavour}

\begin{figure}[tb!]
\centering
  \begin{tikzpicture}
    \begin{feynman}
      \vertex (a) {\(\ell_\alpha\)};
      \vertex [right=1.5cm of a] (b); 
      \vertex [right=4cm of b] (c);
      \vertex [right=1.5cm of c] (d) {\(\ell_\beta\)};
      \vertex [above right=1.8cm and 2cm of b] (top);
      \vertex [above=1.5cm of top] (gamma) {\(\gamma\)};

      \diagram* {
        (a) -- [fermion, thick] (b) -- [fermion, thick, edge label=\(\nu_i\)] (c) -- [fermion, thick] (d),
        (b) -- [boson, half left, looseness=1.5, edge label'=\(W\), midway] (c),
        (top) -- [photon] (gamma),
      };
    \end{feynman}
  \end{tikzpicture}
\caption{Feynman diagram relevant for the muon decay $\mu \to e \gamma$. The fermion mediators include the light and heavy neutral fermions.}
\label{fig:muegamma}
\end{figure}
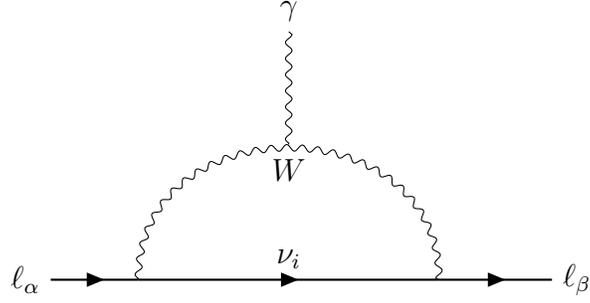

A typical signal of many BSM models, and in particular low-scale seesaws, is the LFV process $\ell_\alpha\to \ell_\beta \, \gamma$. In the context of the Dirac Type-I seesaw family, this process is generated at the 1-loop level and mediated by the $W$ boson and the neutral fermions, see Fig.~\ref{fig:muegamma}. The loop calculation is well known, see for example~\cite{Lavoura:2003xp}, and leads to the following branching ratio
\begin{equation}
    \text{BR}(\ell_\alpha \to \ell_\beta \, \gamma) = \frac{\alpha_W^3 s_W^2}{1024 \pi^2}\left(\frac{m_{\ell_\alpha}}{m_W}\right)^4 \frac{m_{\ell_\alpha}}{\Gamma_{\ell_\alpha}} \left|\left(M_1 M_N^{-1} (M_N^{-1})^\dagger M_1^\dagger\right)_{\alpha \beta}\right|^2 \, .    \label{eq:muegammageneral}
\end{equation}
The model-dependence is encoded in the matrices $M_1$ and $M_F$, while $\alpha_W$, $s_W^2$, $m_{\ell_\alpha}$, $m_W$ and $\Gamma_{\ell_\alpha}$ are the standard electroweak parameters. In a given model, we can compare Eq.~\eqref{eq:muegammageneral} with the experimental constraints, as we will do in Sect.~\ref{sec:pheno}. In particular, the most explored and promising leptonic decay is $\mu \rightarrow e \gamma$; for this particular process the MEG II collaboration~\cite{MEGII:2021fah} has reported the limit~\cite{MEGII:2025gzr}
\begin{equation} \label{eq:meg}
    \BR\left(\mu \to e \, \gamma \right)\lesssim 1.5 \, \cdot \, 10^{-13} \hspace{0.5cm} \text{MEG II (current)} .
\end{equation}
Which the collaboration expects to improve to
\begin{equation} \label{eq:megii}
        \BR\left(\mu \to e \, \gamma \right) \lesssim 6 \, \cdot \, 10^{-14} \, \hspace{0.5cm} \text{MEG II (future)} .
\end{equation}
Although a rigorous numerical analysis will be performed in each specific model, let us make some approximations to derive a naive estimate for the mass parameters required in a model to obtain a sizable $\BR\left(\mu \to e \, \gamma \right)$. Given a model, we can always choose to work in a basis where both the charged lepton mass matrix and $M_N$ are diagonal and real positive matrices. In this basis, and using the parametrization given in Eq.~\eqref{eq:parametrization}, we can express the branching ratio as
\begin{equation}
\text{BR}(\ell_\alpha \to \ell_\beta \,  \gamma) = \frac{\alpha_W^3 s_W^2}{1024 \pi^2}\left(\frac{m_{\ell_\alpha}}{m_W}\right)^4 \frac{m_{\ell_\alpha}}{\Gamma_{\ell_\alpha}} \left|\left(M_\nu M_2^{-1} (M_2^\dagger)^{-1} M_\nu^\dagger\right)_{\alpha \beta}\right|^2 \, . \label{eq:muegammaparam} 
\end{equation}
Furthermore, if we assume that $M_N$ is degenerate, $M_N \equiv m_n \, \id_N$, we can without loss of generality choose to work with $M_2$ diagonal and real positive. Additionally, consider the limit where $M_2$ is also degenerate, $M_2 \equiv m_2 \, \id_N$~\footnote{\label{footnote}From now on, we will use capital letters for mass matrices, while lowercase letters will be reserved for the degenerate cases.}. We call this simplifying limit \textit{the degenerate BSM limit}, which will be useful for obtaining analytic results in a simple way. The branching ratio is now found as
\begin{equation} 
\text{BR}(\ell_\alpha \to \ell_\beta \,  \gamma) = \frac{\alpha_W^3 s_W^2}{1024 \pi^2}\left(\frac{m_{\ell_\alpha}}{m_W}\right)^4 \frac{m_{\ell_\alpha}}{\Gamma_{\ell_\alpha}} \frac{\left|\left(M_\nu M_\nu^\dagger\right)_{\alpha \beta} \right|^2}{m_2^4}. \label{eq:muegammadegenerate} 
\end{equation}
Note that the quantity $|\left(M_\nu M_\nu^\dagger\right)_{\alpha \beta}|^2$ with $\alpha \neq \beta$ does not depend on the mass of the lightest neutrino, only on the squared mass differences and the mixing angles. Therefore, in this simplified scenario, the combination $(m_2)^4 \, \text{BR}(\ell_\alpha \to \ell_\beta \,  \gamma)$ is an experimentally determined quantity. In the $\mu \rightarrow e \, \gamma$ case, assuming the best-fit values for the SM and neutrino parameters as well as normal ordering, we find
\begin{equation} 
m_2^4 \, \BR(\mu \to e \gamma) =\frac{\alpha_W^3 s_W^2}{1024 \pi^2}\left(\frac{m_\mu}{m_W}\right)^4 \frac{m_\mu}{\Gamma_\mu} \left|\left(M_\nu M_\nu^\dagger\right)_{21} \right|^2 \approx 6.32 \times 10^{-11} \text{ eV}^4\, .
\end{equation}
On the other hand, using the current limits on the $\mu \to e \, \gamma$ branching ratio, we can derive a maximum value for the $m_2$ mass scale, imposing a limit on how low the low-scale of the model can be,
\begin{equation} \label{eq:M2max}
    m_2 \geq 4.5 \text{ eV (MEG II limit)} \, , \hspace{1cm}  m_2 \geq 5.7 \text{ eV (MEG II future limit)} \, .
\end{equation}
While this approximate value does not strictly hold in the non-degenerate scenario, the overall qualitative behavior predicted by the analytical approximation is still well captured.

\begin{figure}[tb!]
  \centering
  \begin{subfigure}{0.42\linewidth}
    \includegraphics[width=\linewidth]{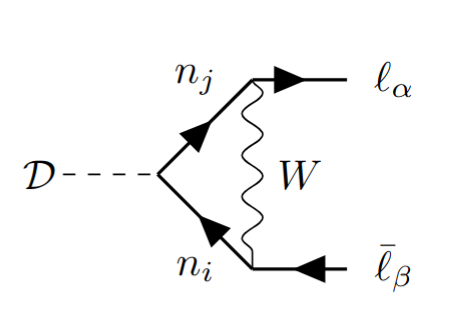}
    \caption{\textbf{$\boldsymbol{W}$ boson contribution}}
    \label{fig:Wdiag}
  \end{subfigure}
  \begin{subfigure}{0.57\linewidth}
    \includegraphics[width=\linewidth]{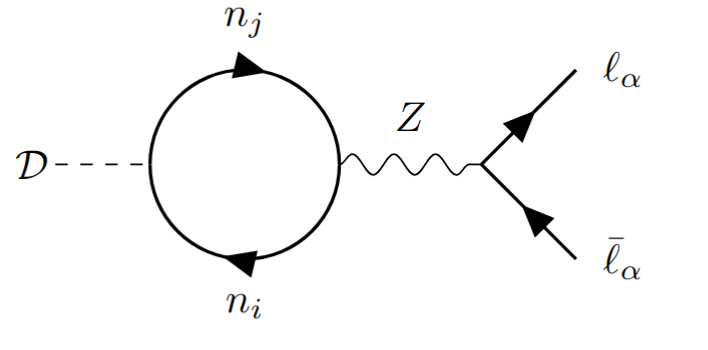}
    \caption{\textbf{$\boldsymbol{Z}$ boson contribution}}
    \label{fig:Zdiag}
  \end{subfigure}
  \caption{Feynman diagrams leading to the 1-loop coupling of the
    Diracon to a pair of charged leptons. \label{fig:mueDdiagrams}}
\end{figure}

On the other hand, the models showcased here also feature a flavour-violating Diracon via the mixing-induced gauge interactions of the neutral fermions, as seen in Fig.~\ref{fig:mueDdiagrams}. The general calculation of the 1-loop coupling of the Diracon to a pair of charged leptons was performed in~\cite{Herrero-Brocal:2023czw}~\footnote{While this reference focused on Majorana neutrino models, and hence on Majoron interactions, all analytical results can be readily adapted for the case of a Diracon.} and we show the details in App.~\ref{app:coupling}. This interaction depends on the tree-level couplings of the Diracon, i.e., it is model dependent. As shown in Sec.~\ref{sec:pheno}, the different $U(1)_D$ breaking patterns in each model will lead to a neutrino-mass suppressed Diracon coupling to charged leptons in the canonical model, while this interaction will not be suppressed by neutrino masses in the enhanced Diracon model. In order to compare the predictions of these models with the experimental constraints, we write the general Diracon-charged lepton effective interactions, given by
\begin{equation} \label{eq:majoronpheno}
\mathcal{L}_{\ell\ell \D} = \D \, \bar{\ell}\left( S_L P_L + S_R P_R\right) \ell + \hc  =\D \, \bar{\ell}\left( S P_L + S^\dagger P_R\right) \ell \, ,
\end{equation}
with $S = S_L + S_R^\dagger$. There are stringent constraints on both diagonal and off-diagonal Diracon couplings to charged leptons~\cite{Escribano:2020wua}. The flavour-conserving couplings are constrained by energy loss mechanisms in astrophysical observations~\cite{DiLuzio:2020wdo,Straniero:2020iyi,Calibbi:2020jvd,Croon:2020lrf, Caputo:2021rux,Caputo:2022rca,Fiorillo:2025sln} and yield
\begin{align}
\label{eq:stellarcooling}
    \left| \text{Im} \left(S_{11}^\text{exp}\right) \right| &< 1.48 \times 10^{-13} \, , \\
    \left| \text{Im} \left(S_{22}^\text{exp}\right) \right| &< 3.1 \times 10^{-9} \, .
    \label{eq:stellarcooling2}
\end{align}
The presence of non-zero off-diagonal couplings in Eq.~\eqref{eq:majoronpheno} allows for the non-standard decay $\mu^+ \to e^+ \, \D$. In particular, we find 
\begin{equation} \label{eq:gamma}
\Gamma(\ell_\alpha \to \ell_\beta \, \D) = \frac{m_{\ell_\alpha}}{32 \, \pi} \, \left| \widetilde S^{\alpha \beta} \right|^2 \, ,
\end{equation}
where we have defined 
\begin{equation}
  \left| \widetilde S^{\alpha \beta} \right| = \left( \left| S^{\alpha \beta }_L \right|^2 + \left| S^{\alpha \beta }_R \right|^2 \right)^{1/2} \, .
  \label{eq:Stilde}
\end{equation}
The best limits on this process were obtained at TRIUMF~\cite{Jodidio:1986mz}. Taking into account all possible chiral structures for the Diracon coupling, one can estimate the bound~\cite{Hirsch:2009ee}
\begin{equation} \label{eq:majoronlimit}
    \BR \left( \mu \to e \, \D \right) \lesssim 10^{-5} \, .
\end{equation}
%
%
%
Future experiments such as Mu3e~\cite{Hesketh:2022wgw,Perrevoort:2024qtc} or COMET~\cite{COMET:2018auw,Xing:2022rob} will improve these constraints. In particular, the expected future sensitivities for a massless Diracon are expected to be
\begin{align}
    \BR\left(\mu \to e \, \D \right)_\text{Mu3e} &\lesssim 6 \,\cdot \, 10^{-7} \, , \\
    \BR\left(\mu \to e \, \D \right)_\text{COMET} &\lesssim 4.6 \,\cdot \, 10^{-9} \, . \label{eq:Comet}
\end{align}
Other muon decays with a Diracon in the final state that can be used to probe a flavour-violating coupling are $\mu \to e \D \gamma$~\cite{Hirsch:2009ee,Jho:2022snj,Herrero-Brocal:2023czw}, $\mu \to eee \D$~\cite{Knapen:2023zgi} and $\mu \to e \D\D$, as well as analogous $\tau$ decay processes. In what follows we will, however, focus on the 'golden' flavour violating signatures $\mu \to e \gamma$ and $\mu \to e \D$.

\section{Cosmology and $N_{\text{eff}}$}
\label{sec:Cosmo}
\subsection{Dirac neutrinos and $\dneff$}

Models that predict new light degrees of freedom are subject to strong cosmological limits. For sufficiently strong interactions with the SM, they will be produced abundantly in the early Universe, and their respective energy density contributes to the total expansion rate as such. Now it has been known for long that cosmological observables test the expansion rate (indirectly) at selected points in the thermal history~\cite{PhysRevLett.43.239,Dolgov:2002wy,RevModPhys.53.1}.
Any additional radiation at the time of BBN will affect the delicate balance of primordial element abundances, and the angular scale of acoustic peaks and the damping tail in the CMB power spectrum are particular sensitive probes of the early expansion rate and the ratio of the radiation- and matter-like energy budgets~\cite{2013PhRvD..87h3008H}.

It is conventional to normalize additional radiation beyond photons to that of a single active neutrino flavour. This effective number of neutrinos $N_\text{eff} = (8/7)\,(11/4)^{4/3} \rho_\text{rad}/\rho_\gamma$ is predicted to be $N_\text{eff} = 3.044$~\cite{deSalas:2016ztq,Froustey:2020mcq,EscuderoAbenza:2020cmq,Akita:2020szl,Bennett:2020zkv} in the SM. Deviations from the above SM expectation are usually parameterized as $\dneff = N_\text{eff}-3.044$.

If a new light degree of freedom is in chemical equilibrium with the SM plasma at some early time, its contribution to $\dneff$ can be obtained from its decoupling temperature $T_\text{dec}$ as
\begin{equation}
    \dneff  \simeq 0.027\,g_x \left(\frac{106.75}{g_\star(T_\text{dec})}\right)^{4/3}\,.
    \label{eq:basis}
\end{equation}
Here $g_x$ denotes the effective internal degrees of freedom, i.e. $g_x = 1$ for scalars, $g_x = 7/8\times 2$ for a Weyl fermion, etc. This highlights as to why cosmology is so powerful in constraining models of Dirac neutrinos -- they necessarily come with at least two additional light Weyl fermions, which upon thermalization above the EW scale would contribute as $\dneff \simeq 0.09$. A full three thermalized right-handed neutrinos, and possibly an additional Nambu-Goldstone boson, would contribute a significant $\dneff \simeq 0.14$ and $\dneff \simeq 0.17$, respectively. However, interactions can be too feeble to establish full thermalization of the light degrees of freedom, and the production proceeds from freeze-in.

Cosmological limits on $\dneff$ already give valuable input for model building, with the long standing Planck limit $\dneff < 0.285$ at $95\%$ C.L.~\cite{Planck:2018vyg,Abazajian:2019oqj}. More recently, the Atacama Cosmology Telescope (ACT)~\cite{ACT:2025tim,ACT:2025fju} pushed the limit to $\dneff<0.17$ at $95\%$ C.L. by including large and medium scale data from Planck. Joint CMB+BBN analyses find $\dneff < 0.180$~\cite{Yeh:2022heq} and $\dneff<0.163$~\cite{Fields:2019pfx} at $95\%$ C.L. each. This highlights how models leading to a Dirac nature of neutrinos are already being tested, with future experiments narrowing the window. Indeed, the Simons Observatory (SO)~\cite{2019BAAS...51g.147L} and SPT-3G~\cite{SPT-3G:2014dbx} anticipate a final sensitivity of $\dneff < 0.12$ at $95\%$C.L.~\cite{Abazajian:2019oqj}. The original proposal of CMB-S4~\cite{Abazajian:2019eic} aims for $\dneff<0.06$, and far-future proposals such as CMB-HD~\cite{Sehgal:2019ewc} intend to to go beyond $\dneff < 0.02$, with the potential to rule out a single thermalized Nambu-Goldstone boson.

\subsection{Boltzmann equations}
We introduce the energy density for the right-handed neutrino $\rho_{\nu_R}$, which we treat as a common species for all right-handed (anti-)neutrinos~\cite{Adshead:2022ovo,Luo:2020fdt,Luo:2020sho}. Similarly, the scalar sector after symmetry breaking is described as an effective species with $g=2$ degrees of freedom if $T\gg m_\mathcal{S}$ and $g=1$ for $T\ll m_\mathcal{S}$. This is justified, since we adopt $\lambda_\sigma \gg \lambda_{H\sigma}$ throughout.

Thus, we augment the thermal history by introducing the two effective radiation like fluids,
\begin{align}
    \frac{d\rho_{\nu_R}}{dt} &= -4H\rho_{\nu_R} + \mathcal{C}^{(\rho)}_{\SM\to\nu_R} + \mathcal{C}^{(\rho)}_{\sigma\to\nu_R}\,,\\
    \frac{d\rho_{\sigma}}{dt} &= -4H\rho_{\sigma} + \mathcal{C}^{(\rho)}_{\SM\to\sigma} - \mathcal{C}^{(\rho)}_{\sigma\to\nu_R}\,,\\
    \frac{d\rho_\SM}{dt} &= -3H(\rho_\SM+P_\SM) - \mathcal{C}^{(\rho)}_{\SM \to \nu_R}- \mathcal{C}^{(\rho)}_{\SM\to\sigma}\,.
\end{align}
Here $\mathcal{C}^{(\rho)}_{x\to y}$ is the integrated collision operator describing the energy transfer from sectors $x\to y$, and $H^2=\frac{8\pi G}{3} (\rho_\SM + \rho_{\nu_R} +\rho_\sigma)$ is the Hubble rate. In general, the collision operator connecting species $x$ and $y$ can be written as the sum of all processes connecting initial states $i$ and final states $f$ that contain at least one $x$, $y$,
\begin{equation}
\begin{aligned}
    \mathcal{C}^{(\rho)}_{x\to y} =& \sum_{i,f \ni x,y} \int d\Pi_{i_1}...d\Pi_{i_n} d\Pi_{f_1}...d\Pi_{f_m}
    (2\pi)^4 \delta^{(4)}\left(\sum p_i -\sum p_f\right) \, (E_{y}-E_x) \\
    \times & \left( \abst{\mathcal{M}_{i\to f}}^2 f_{i_1}...f_{i_n}\bar f_{f_1}...\bar f_{f_m}  - \abst{\mathcal{M}_{f\to i}}^2  f_{f_1}...f_{f_m}\bar f_{i_1}...\bar f_{i_n} \right)\,.
    \label{eq:coll_rho}
\end{aligned}
\end{equation}
Here $d\Pi_i = \frac{d^3p_i}{(2\pi)^3 2E_i}$ is the phase space element, $f_x=f_x(E_x)$ is the phase space density, and $\bar f_x = 1\pm f_x$ is the final state phase space density, where the sign depends on whether the final state is a boson (fermion).
We also assume $\abst{\mathcal{M}_{i\to f}}^2 = \abst{\mathcal{M}_{f\to i}}^2$ for simplicity. In our convention, the matrix element includes symmetry factors, i.e. $1/n!$ for $n$ identical particles in the initial and final states~\cite{Kolb:1990vq}.

We include all $2\leftrightarrow2$ and $1\leftrightarrow2$ processes that can be constructed from the new scalar and Yukawa terms, which we list in Appendix \ref{app:scalar} and \ref{app:Yukawa}. Moreover, we include processes between the radial scalar and other SM particles that arise from the mixing of Higgs and radial scalar.

As far as the SM is concerned, we assume that the equilibrium assumption is never violated. Thus, we can write
\begin{equation}
    \rho_\SM=\frac{\pi^2}{30}g_{(\rho)}T^4\,,\quad P_\SM=\frac{\pi^2}{90}g_{(P)}T^4\,,
\end{equation}
to infer a relation between energy density and temperature, and prescribe Fermi-Dirac and Bose-Einstein distributions in the collision integral. For the possibly non-equilibrated effective $\nu_R$ and $\sigma$ species, we use equilibrium distributions, which are excellent approximations for the freeze-in regime as well as the fully coupled limit and serve as a smooth interpolation between the two regimes~\cite{Luo:2020fdt,Adshead:2022ovo,Herbermann:2025uqz}.

We use the numerical implementation developed in~\cite{Herbermann:2025uqz,Luo:2020fdt} that uses Monte Carlo integration to find an exact numerical result for the collision integral. The scheme thus includes the full relativistic effects and final state statistics. We begin integration at some high temperature $T_i$ with negligible initial abundances for the new particles, and assume full thermal equilibrium for the SM. We typically choose $T_i \sim 10^3 \Lambda$, where $\Lambda$ is a placeholder for the highest scale present in the theory (usually $M_N$ or $v_\sigma$). This allows the system to fully relax into an equilibrium state if interactions are sufficiently strong or, in the case of freeze-in-like production, the error we introduce for non-UV dominated freeze-in processes is negligible. We integrate the system of equations forward in time up to $T_\SM \sim \mathrm{\,MeV}$.

The models we consider allow for unstable particles in the s-channel, and the mediator may also be in thermal equilibrium at the same time, thus the resonant on-shell creation of a mediator and its subsequent decay are physically equivalent to the decay of the mediator in thermal equilibrium. When both processes are taken into account, issues of double counting arise~\cite{Weldon:1983jn,Bringmann:2021sth,DeRomeri:2020wng}.  We address the resonances in the narrow width approximation (NWA), so we adopt the Breit-Wigner prescription for the squared propagator close to the resonance.
The matrix element for on-shell production and subsequent decay can be written as a product of the matrix elements of inverse decay and decay of the mediator, i.e. we formally write
\begin{equation}
\abst{\mathcal{M}_{i \to f}}^2 = \frac{\abst{\mathcal{M}_{i \to \mathrm{mediator}}}^2\abst{\mathcal{M}_{\mathrm{mediator}\to f}}^2}{M\Gamma} \, \pi \, \delta(s-M^2)\,.
\end{equation}
The decomposition is exact for scalar mediators on resonance and correct up to spin correlations for fermions or vector bosons on resonance~\cite{Bringmann:2021sth,Bringmann:2017sko,Uhlemann:2008pm,Herbermann:2025uqz}, but appropriate generalizations for fermionic and vector mediators exist~\cite{Uhlemann:2008pm}.

Away from the resonance, we take the $2\to 2$ amplitude, but cut the resonant region. For the resonant contribution, we either consider the inverse decay and subsequent decay for an out of equilibrium mediator, or if the mediator is in equilibrium, we consider the decay from equilibrium instead. Thus, we avoid the double counting issue in a conservative manner. The approach is reminiscent to that in~\cite{DeRomeri:2020wng}.

\subsection{Phase transitions and thermal corrections}
The leading effect of thermal corrections can be approximated by including thermal masses. The effect on the end result is expected to be minimal (see also~\cite{Herbermann:2025uqz}), but is straightforwardly included in the numerical scheme.
Finite temperature correction of the scalar potential however play a crucial role due to symmetry breaking, as this determines the spectrum and interactions of the theory at any given temperature.

We generically assume a regime of weak portal couplings $\lambda_{H\sigma}\ll \lambda_\sigma$. Thus, we assume that the two sectors are only weakly coupled, and the phase transitions proceed independently of each other. For the SM, we justify this by means of portal suppressed corrections and so it proceeds as SM-like at $T^c_{H}\simeq 160\mathrm{\,GeV}$. The $\sigma$-transition is somewhat different, and Higgs related corrections can be dominant even for the small portal depending on the new scalar scale. As we checked explicitly, the results we obtain are only subdominantly affected by the detailed timing of the transition. The dominant effect results from the presence or absence of certain interaction vertices at their respective production peak, in particular decay modes that peak at $T\lesssim m$. This is corroborated by similar conclusions found in~\cite{DeRomeri:2020wng}.

Thus, considering the thermally corrected mass terms we prescribe (e.g.~\cite{Katz:2014bha,Matsedonskyi:2020mlz})
\begin{equation}
\begin{aligned}
	-\mu_H^2(T) &\simeq -\mu_H^2 + \left(\frac{3}{16}g^2+\frac{1}{16} g^{\prime 2} + \frac{1}{4}y^2_t + \frac{1}{2}\lambda \right) T^2\,\\
	-\mu_\sigma^2(T,T_\sigma) &\simeq -\mu_\sigma^2 + \frac{1}{3}\lambda_\sigma T_\sigma^2 + \frac{1}{6}\lambda_{H\sigma} T^2\,.
    \label{eq:thermal_corrections_scalar}
\end{aligned}
\end{equation}
Here, $g,g^\prime$ are the couplings to $W$ and $Z$ bosons, $y_t$ is the top Yukawa and $\lambda$ the Higgs quartic scalar coupling. There is also a correction to the $\sigma$ related terms of the potential that arises from the portal coupling. We define the temperature of the $\sigma$-transition $T^c_{\sigma}$ by equating absolute values of thermal corrections and the vacuum term in Eq.~\eqref{eq:thermal_corrections_scalar}.

We note that the energy released in the $\sigma$-transition is of the order $\Delta V \sim \lambda_\sigma v_\sigma^4$. The type of the transition determines the fraction of energy that is actually transferred to the hidden sector. Here we stay agnostic about the transition itself, noting that the most energy that could be transferred is the total released energy.
If the hidden sector is coupled to the SM, or the transition occurs before the sectors will couple at a later time, any effect of the transition is washed out. If, however, the transition occurs when the sectors are decoupled, the energy is predominantly released into the decoupled hidden sector and increasing its energy density. Thus, the transition could pose an extra correction to $\dneff$. However, comparing $\Delta V$ to the energy density of $\sigma$ and the SM at $T^c_{\sigma}$ shows that the extra contribution to $\dneff$ is subdominant if the sector saturates its equilibrium abundance unless the thermal corrections in Eq.~\eqref{eq:thermal_corrections_scalar} cancel due to relative signs in couplings.

\section{Phenomenological analysis}
\label{sec:pheno}

We will now study the phenomenology of the canonical and enhanced Diracon models. In both models, which we study separately, we will explore the complementarity between lepton flavour violation and cosmological probes.

\subsection{Model I: Canonical model}
\label{sec:PhenoCanonical}

As mentioned in Sec.~\ref{sec:Flavour}, one of the most promising signals in low-scale seesaws arise from flavour-violating processes. In this work we will focus on the simplest processes in our model, $\ell_\alpha \to \ell_\beta \, \gamma$ and $\ell_\alpha \to \ell_\beta \, \D$.  The general formula for the $\ell_\alpha \to \ell_\beta \, \gamma$ rates is given in Eq.~\eqref{eq:muegammageneral}. In the case of $\ell_\alpha \to \ell_\beta \, \D$, which includes the Diracon in the final state, although the general formula can be computed from the expressions given in Appendix~\ref{app:coupling}, it is illustrative to work in the degenerate BSM simplifying limit discussed in Sec.~\ref{sec:Flavour}, where $M_N = m_N \, \id_2$ and $M_2 = Y' v_\sigma/\sqrt{2} = y' \id_2 v_\sigma/\sqrt{2}$. Under this assumption, the effective Diracon-charged lepton interaction Lagrangian is given by
\begin{equation} \label{eq:degenerateD-cl}
\mathcal{L}_{\ell\ell \D} = - \frac{i \, \D}{96\pi^2 }\frac{y'^2 v_\sigma}{m_N^2} \bar{\ell} \, \left[ M_\ell  \, \textup{Tr}(Y Y^\dagger) \, \gamma_5 +5 \, M_\ell\, Y Y^\dagger \, P_L - 5 \,  Y  Y^\dagger \, M_\ell \, P_R \right] \ell \, .
\end{equation}
While we do not show the non-degenerate expression due to the lengthy and not illuminating formulas involved, note that the Diracon interactions in Eq.~\eqref{eq:degenerateD-cl} are neutrino mass suppressed. Indeed, in the limits $M \to \infty$ or $y' \to 0$ we would get both $m_\nu \to 0$ and $\text{Br}(\mu \to e \D) \to 0$. There is no limit where $m_\nu \to 0$ while keeping a finite term in Eq.~\eqref{eq:degenerateD-cl}.

On the other hand, in the degenerate BSM limit described in Sec.~\ref{sec:Flavour}, we can write the rate of the $\mu \to e \gamma$ radiative decay. Reading Eq.~\eqref{eq:muegammadegenerate}, we find in this model
\begin{align} \label{eq:muegammacanonical}
    \text{BR}(\mu \to e \, \gamma) = 
    \frac{\alpha_W^3 s_W^2}{1024\,  \pi^2}\left(\frac{m_{\mu }}{m_W}\right)^4 \frac{m_{\mu }}{\Gamma_{\mu }} \left(\frac{\sqrt{2}}{v_\sigma y'}\right)^4 \left|\left(U_\ell \widehat{M_\nu}^2 U_\ell^\dagger\right)_{21}\right|^2 \, ,
\end{align}
Where $\widehat{M_\nu}^2 \equiv \widehat{M_\nu} \widehat{M_\nu}^T$ is a $3\times 3$ diagonal matrix of rank $2$. We observe that the $\mu \to e \gamma$ rate is enhanced when $y' v_\sigma \ll v$, in agreement with the qualitative discussion around Eq.~\eqref{eq:lowscalegeneral}. Indeed, the limit $y' \to 0$ implies $m_\nu \to 0$, but the rate for $\mu \to e \gamma$ would still be finite\footnote{Let us note that in Eq. \eqref{eq:muegammacanonical}, we are using the parametrization given by Eq. \eqref{eq:CasasIbarraCanonical}, which is no longer valid in the limit $y' \to 0$. However, this statement remains clearly true when read from Eq. \eqref{eq:muegammageneral}.}. This is again a well-known fact in traditional low-scale seesaw models~\cite{Bernabeu:1987gr}. We can read the experimental constraint on the model from Eq.~\eqref{eq:M2max} just replacing $m_2 = y' v_\sigma / \sqrt{2}$. Comparing the rates of both types of BSM lepton decays in this simplifying limit we find
\begin{equation}
   \frac{\text{BR}(\ell_\alpha \to \ell_\beta \, \D)} {\text{BR}(\ell_\alpha \to \ell_\beta\, \gamma)} \approx \left(\frac{5}{12 \pi^{3/2} \alpha_w^{3/2} s_w} \right)^2  \left(\frac{M_W^2}{v \, m_{\ell_\alpha}} \right)^2 \,  \frac{y'^2}{2} \frac{m_2^2}{v^2} \, .
   \label{eq:ratiocanonical}
\end{equation}
In the case of $(\alpha, \beta)=(\mu, e)$ and assuming that $\mu\to e \gamma$ is within the reach of MEG II (which implies $m_2 < 5.7$ eV, see Eq.~\eqref{eq:M2max}), this is equivalent to
\begin{equation}
    \frac{\text{BR}(\mu \to e \, \D)}{\text{BR}(\mu \to e\, \gamma)} \lesssim 1.02 \cdot 10^{-14} \, y'^2 \, .
\end{equation}
So that if $\mu \to e \gamma$ is observable, then $\mu \to e \D$ is not. On the other hand, in order to increase the ratio of Eq.~\eqref{eq:ratiocanonical} we would need to increase $m_2 = \frac{v_\sigma}{\sqrt{2}} y'$, which from the neutrino mass formula in Eq.~\eqref{eq:Mnucanonical} implies reducing $Y$ or increasing $M_N$, further decreasing both rates. It becomes clear that, in the canonical model, the most promising flavour-violating signal will be $\mu \to e \gamma$, similar to traditional low-scale seesaw models, due to the neutrino mass suppression of the Diracon interactions in this model. 

\begin{figure}
    \centering
    \includegraphics[width=0.45\linewidth]{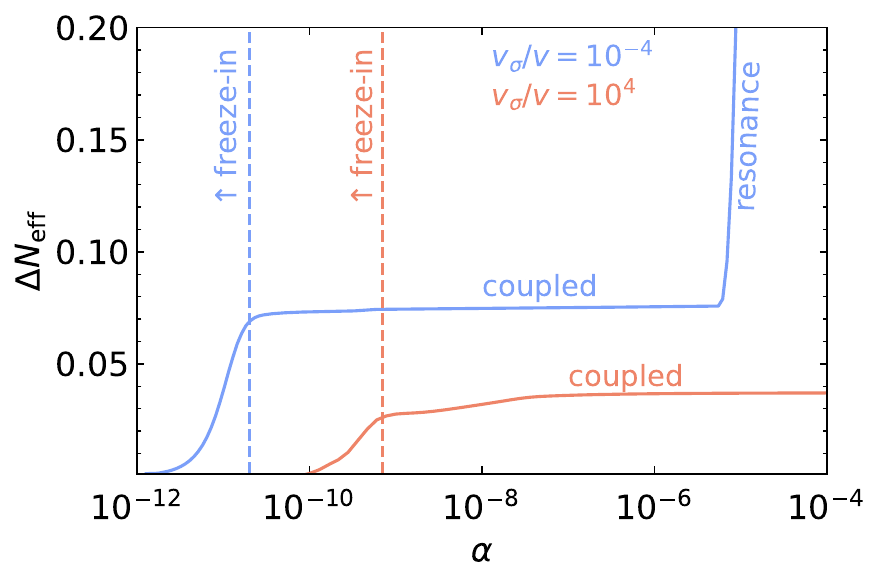}
    \includegraphics[width=0.45\linewidth]{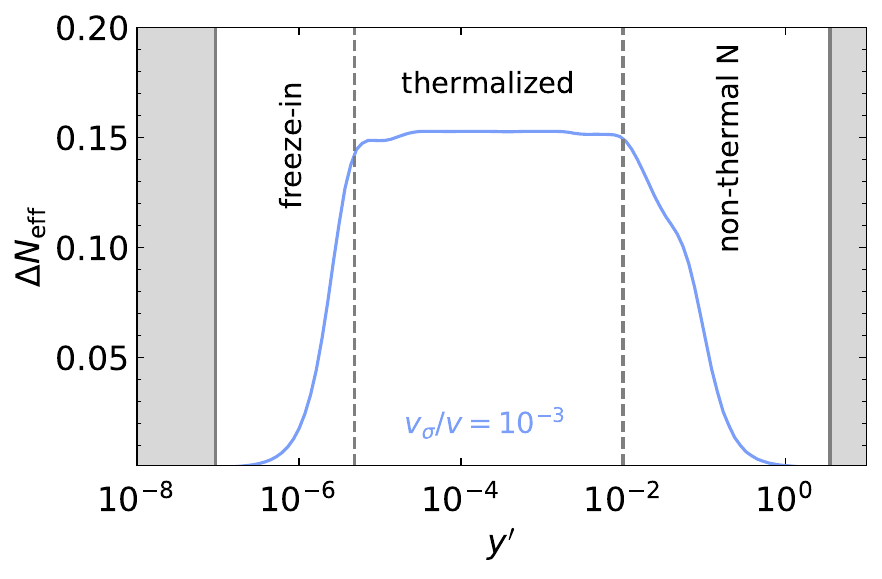}
    \caption{Left: Scalar sector for $v_\sigma/v = 10^{-4}$ and $v_\sigma/v=10^{4}$ respectively. For very small mixing angle $\alpha$, the portal coupling at fixed $v_\sigma$ is suppressed and production proceeds via freeze-in. At larger mixing, the extended scalar spectrum is tightly coupled to the SM. The difference between low and high-scale realizations comes from thermalization of the radial mode. Low-scale realizations also have the possibility of resonant production of Diracons. Right: Fermionic sector with $m_N=1\mathrm{\,TeV}$ and $v_\sigma/v=10^{-3}$. Grey shaded regions are excluded from perturbativity of $Y$ and $Y^\prime$ respectively. For small $y^\prime$, the heavy fermions are thermalized and light degrees of freedom are populated from freeze-in. At larger coupling, $\nu_R$ and $\mathcal{D}$ are coupled to the SM. At even larger values of $y^\prime$, the heavy fermions fall out of equilibrium and production is again suppressed.}
    \label{fig:BP_dneff}
\end{figure}

We will now discuss how these constraints compare to the cosmological ones. To reduce the number of free parameters that would be needed to be scanned over in a full analysis, we adopt the previously discussed limit of degenerate BSM states, but discuss possible modifications in the general case when important.

It is convenient to consider two limiting cases, in which either the scalar sector or the new Yukawa sector dominate. We illustrate the general behavior of the two limits for benchmark points in Fig.~\ref{fig:BP_dneff}. In the scalar limit and for very small mixing/portal coupling, neither the Diracon, nor the radial mode for low-scale realizations thermalize. In this regime, the relic energy density is determined by a freeze-in type production. As we transition to the coupled regime, a moderate increase in $\dneff$ contributions for increasing $\alpha$ is observed from a decreased decoupling temperature. Eventually the contribution flattens out for the high-scale variant, as even larger couplings cannot compensate for the increasingly feeble interactions. For low-scale variations, we can remain coupled to the SM for longer, so the contribution does not necessarily flatten out. Especially in the regime $v_\sigma/v \sim 10^{-2}-10^{-3}$ the light radial mode can be resonantly produced, which leads to stronger limits.

\begin{figure}
    \centering
    \includegraphics[width=0.75\linewidth]{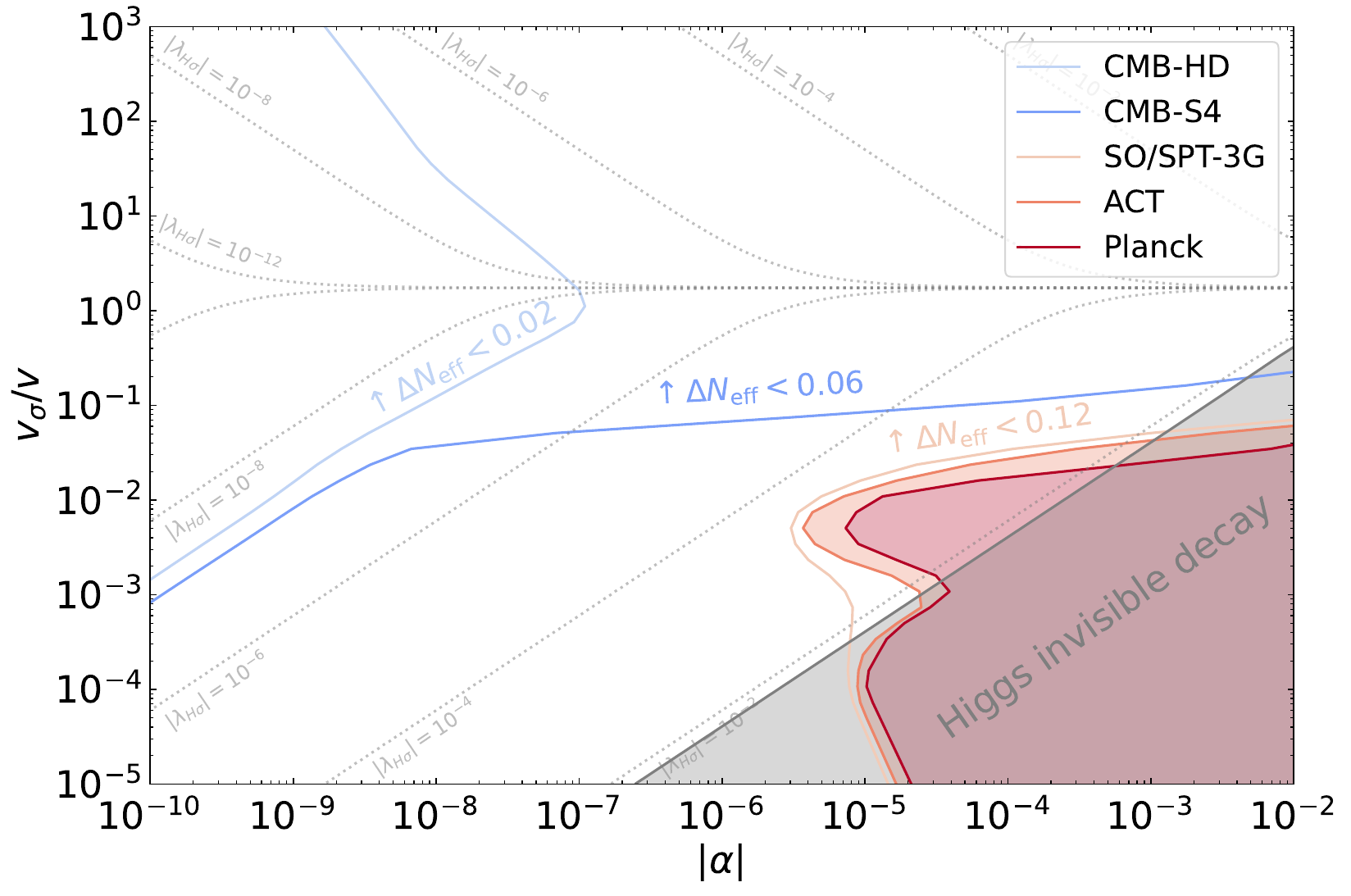}
    \caption{Limits on the scalar sector without considering contributions from the new heavy fermions. Current limits Planck and ACT are weaker than limits from Higgs invisible decay except for the resonant low-scale regime. Future experiments like CMB-S4 will put strong limits on low-scale variations, whereas futuristic proposals like CMB-HD have the potential to constrain the scalar sector on all scales. The scalar self coupling is fixed to $\lambda_\sigma=0.1$, and we show contours of $\vert\lambda_{H\sigma}\vert$ for comparison.}
    \label{fig:scalar_limits}
\end{figure}

In the Yukawa limit, we assume $\alpha=0$ and therefore Diracons and neutrinos are produced through terms involving the heavy new fermions only. Here, the seesaw condition can be understood as fixing the product of the Yukawa couplings, therefore inducing a seesaw relation between $Y$ and $Y^\prime$. For fixed fermion masses, correct neutrino masses and fixed scalar scale, both couplings are subject to perturbativity constraints which we indicate in Fig. \ref{fig:BP_dneff}. Contributions to $\dneff$ can be divided into three regimes. For very small $y^\prime$, we have thermalization of the heavy fermions and population of the light sector via freeze-in. For larger $y^\prime$, we transition into the regime where $\nu_R$ and $\mathcal{D}$ thermalize with the SM through $N$. Eventually $Y$ is too suppressed and $N$ falls out of equilibrium, largely cutting the connection between light degrees of freedom and the SM. The width of the window for efficient production scales as $Y Y^\prime \propto v/v_\sigma$ due to the seesaw condition.

These considerations are useful for understanding the general shape of the limits we find. We show predicted corrections to $\dneff$ for the scalar sector in Fig. \ref{fig:scalar_limits}. Experiments like CMB-S4 will put sizeable limits on the low-scale variant, as for values of $\dneff = 0.06$ a fully thermalized Diracon and radial mode below the electroweak scale is in reach. The asymptotic limit for small mixing follows from the dependence between mixing and the portal coupling -- indeed the decoupling process and hence the asymptotic limit for small mixing is determined by the portal coupling. For the even smaller benchmark $\dneff = 0.02$, we become sensitive to a single scalar degree of freedom even if not fully thermalized above the electroweak scale.

\begin{figure}
    \centering
    \includegraphics[width=0.75\linewidth]{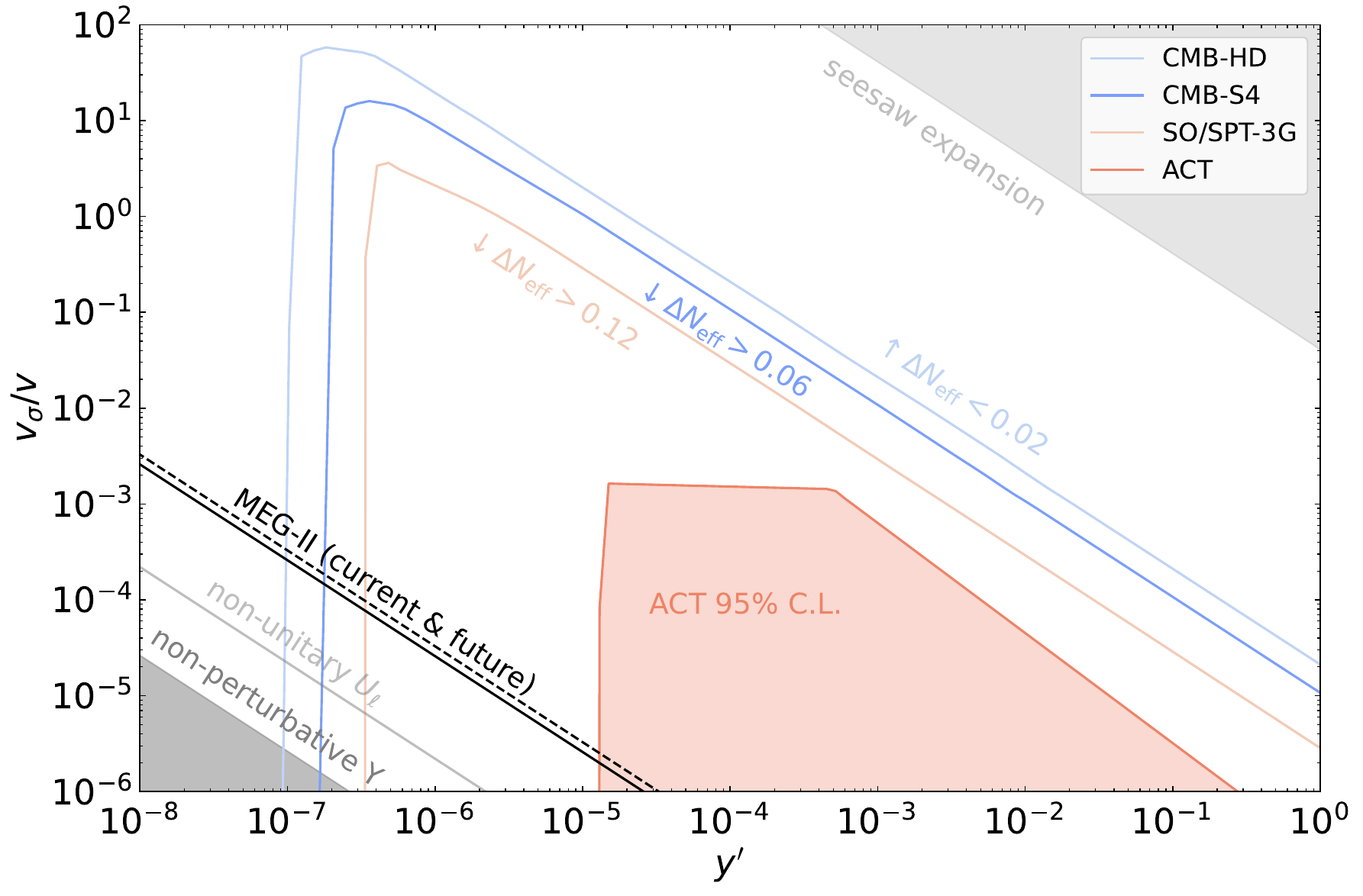}
    \caption{Limits in the canonical model coming from the new fermionic sector in the limit $\alpha=0$ and a benchmark value of $m_N=1\mathrm{\,TeV}$ in the degenerate BSM approximation discussed in the main text. The shape of the constraints is a manifestation of the seesaw relation between Yukawas. The limits from ACT arise from right-handed neutrinos and the Diracon being coupled to the SM below the electroweak scale. We indicate also limits from MEG II, as well as the regions which would lead to non-perturbative $Y$ or not satisfy the seesaw expansion condition $M_2 \ll M_N$ which allows the perturbative diagonalization of the mass matrix.}
    \label{fig:fermion_1TeV}
\end{figure}

\begin{figure}
    \centering
    \includegraphics[width=0.49\linewidth]{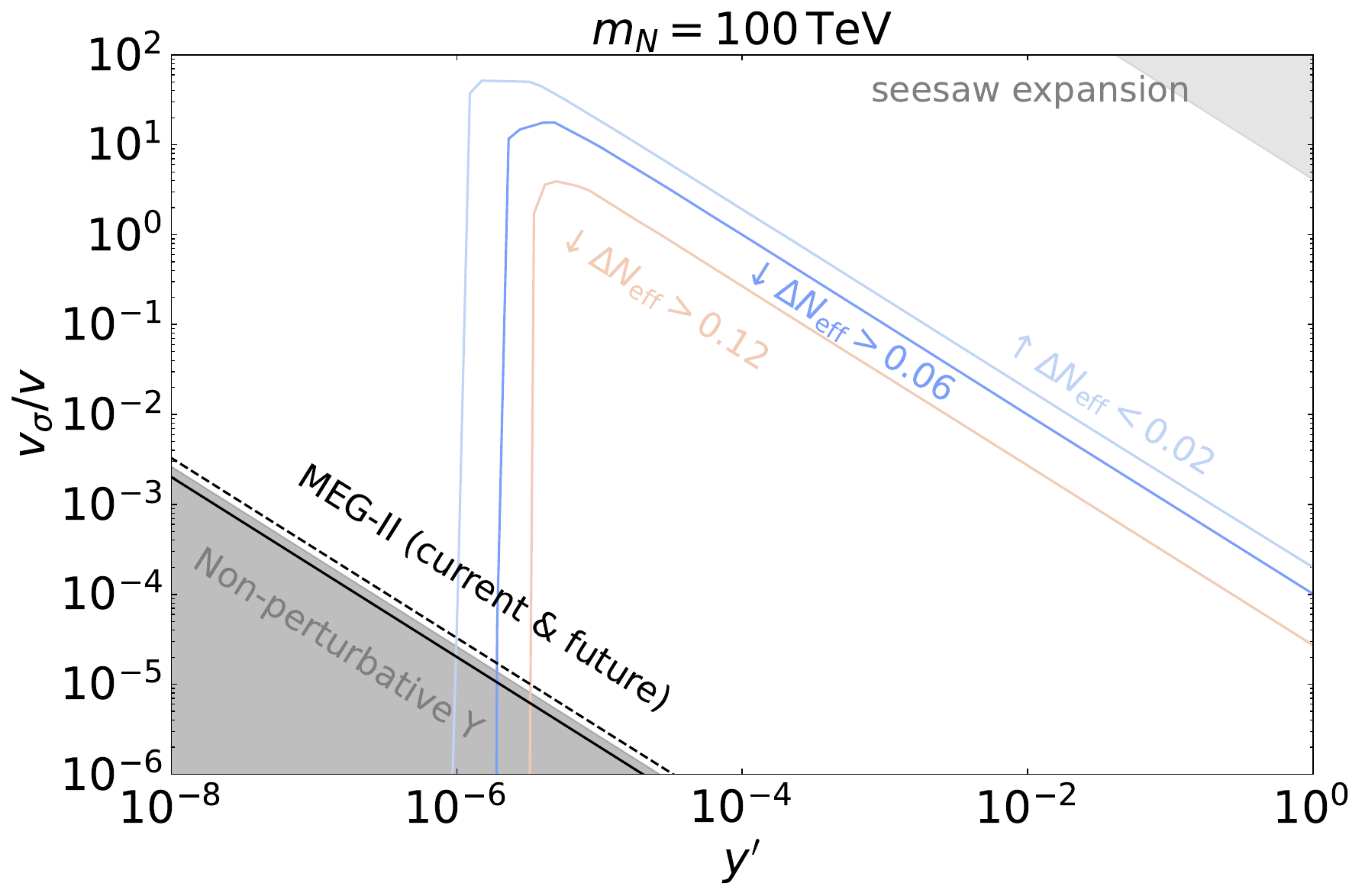}
    \includegraphics[width=0.49\linewidth]{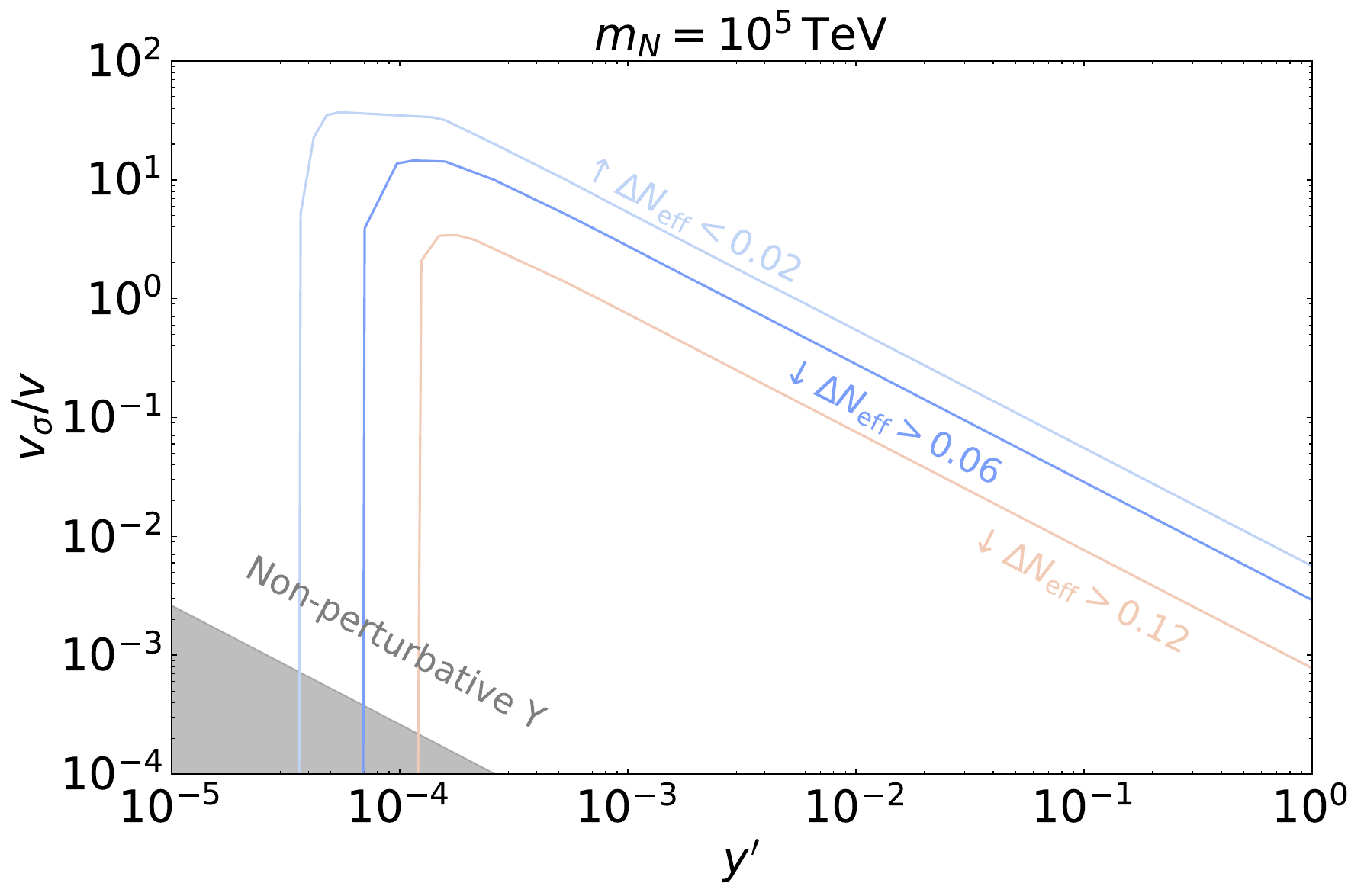}
    \includegraphics[width=0.49\linewidth]{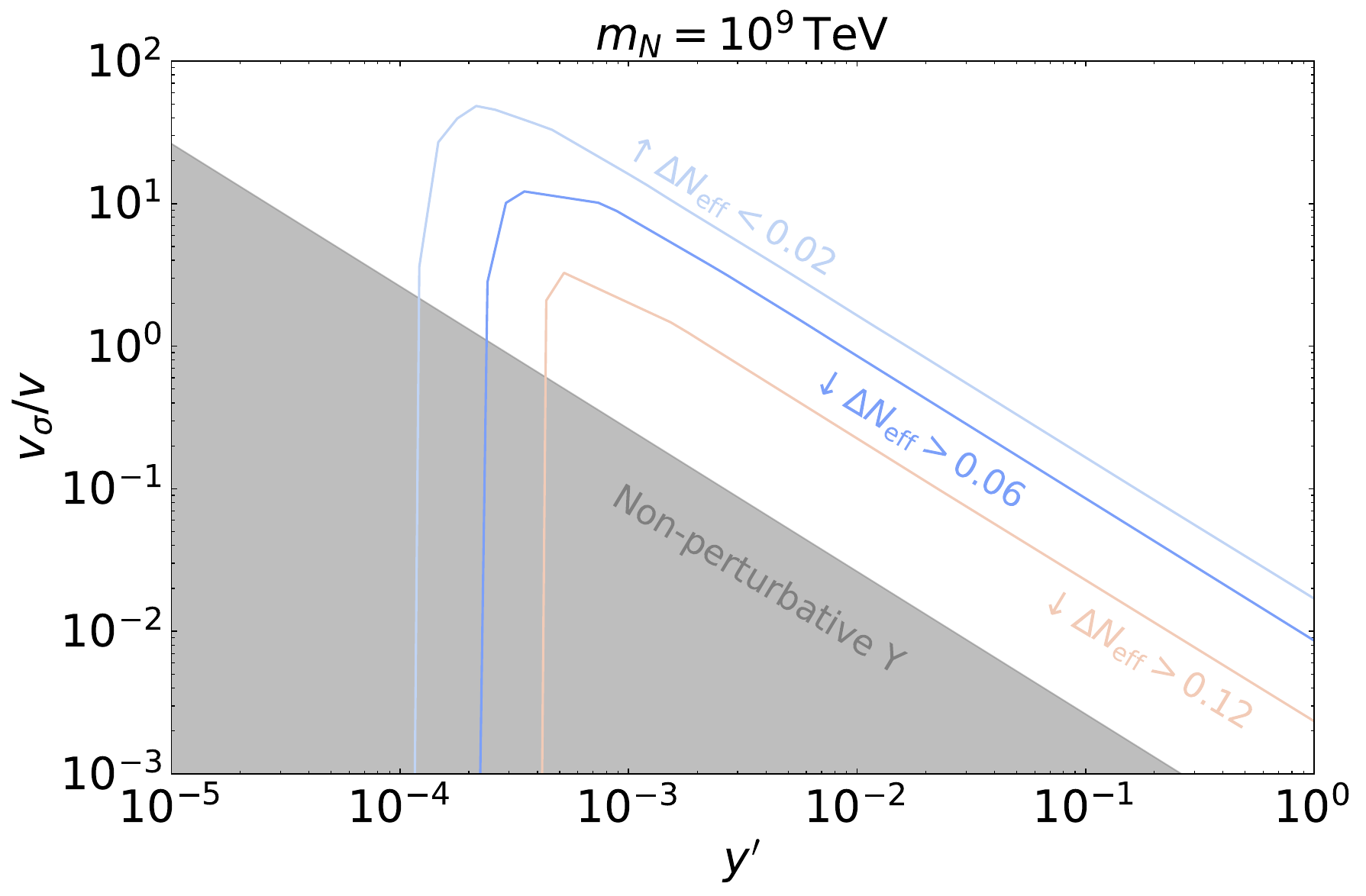}
    \caption{Limits in the canonical model for heavy new fermions in the limit of negligible scalar sector contributions to $\dneff$. We present results for representative benchmark values $m_N=100\mathrm{\,TeV},\, 10^5\mathrm{\,TeV}$ and $10^{9}\mathrm{\,TeV}$, again in the degenerate BSM limit. The shaded region indicates non-perturbative $Y$. The results highlight the potential of future CMB surveys to tightly constrain the fermionic sector up to very high scales.}
    \label{fig:fermion_other}
\end{figure}

The shape of exclusion regions in Figs.~\ref{fig:fermion_1TeV} and \ref{fig:fermion_other} follow directly from the three production regimes we identify and are a manifestation of the seesaw mechanism. When applicable, cosmological limits tend to be stronger than flavour constrains, however, the latter dominate in regions that cosmology cannot test. For the case of $m_N = 1\mathrm{\,TeV}$, we note that we have current limits from ACT from thermalization of Diracons and neutrinos below the EW scale. For larger masses they are absent and constraints shift to larger values of $y^\prime$ in general, which again is a manifestation of the seesaw condition and the dependence of the production amplitudes on the Yukawa couplings.

\begin{figure}
    \centering
    \includegraphics[width=0.49\linewidth]{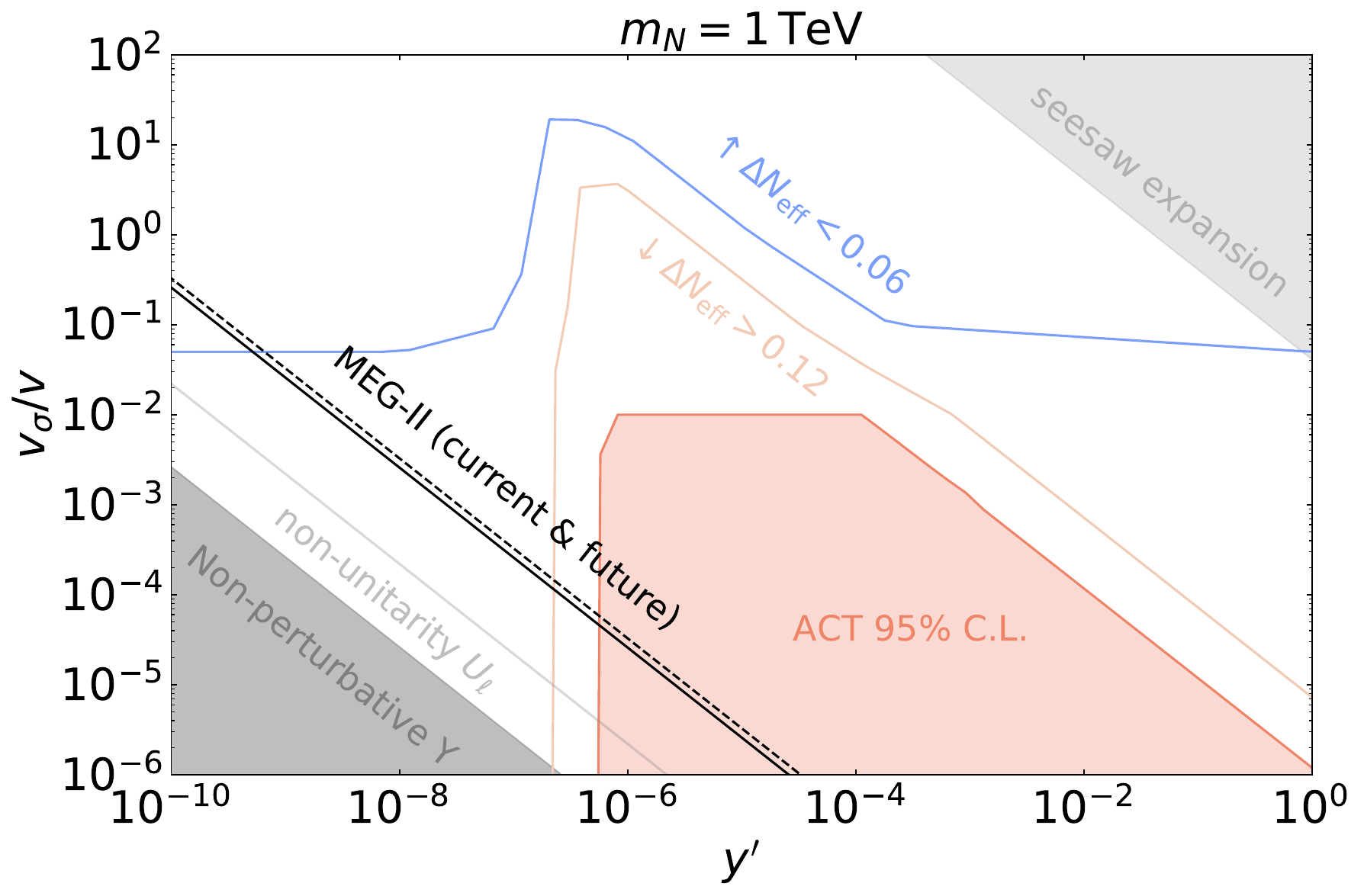}
    \includegraphics[width=0.49\linewidth]{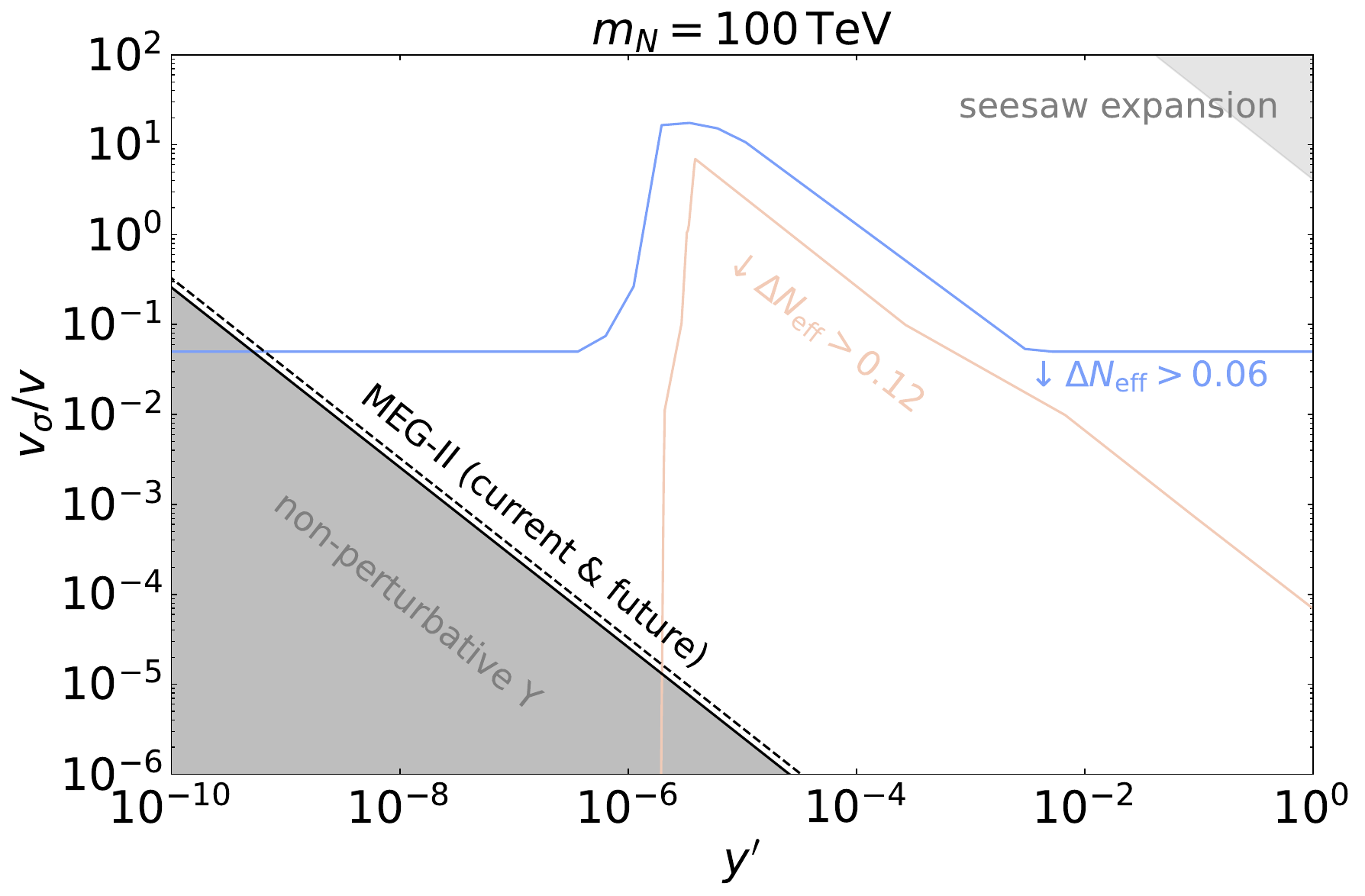}
    \caption{Results of combined full scans at mixing $\alpha=10^{-6}$ for $m_N=1\mathrm{\,TeV}$ (left) and $100\mathrm{\,TeV}$ (right). For larger values of $\dneff$, limits are slightly stronger when compared to the fermionic cases only. For smaller values, here represented by $\dneff=0.06$, constraints flatten out as the scalar contribution alone probes this values in the low-scale regime. There is no limit for CMB-HD, as the scalar contribution alone would be rule out on all scales shown in the figures.}
    \label{fig:full_scans}
\end{figure}

Understanding the combined result is straightforward, and we show selected results for a benchmark of $\alpha = 10^{-6}$ in Fig. \ref{fig:full_scans}. In the regime where limits are strongest, the addition of the scalar sector makes limits moderately stronger, as we can keep the Goldstone coupled to the SM at lower temperatures. In regions where $\nu_R$ production is suppressed, limits are approximately given by the scalar sector only. Thus, to good approximation, it is sufficient to study the sectors independently and apply the relevant individual limits for any chosen parameter configuration.

In principle, high scales for the scalar and the new fermions can be tested, as is evident from Fig. \ref{fig:scalar_limits} and \ref{fig:full_scans}. Courtesy of the Dirac nature of neutrinos and the spontaneous breaking of an exact global $U(1)$, light degrees of freedom are available at all scales. The caveat here is that we implicitly make the assumption of reheating temperatures $T_\text{reh}\sim 10^2-10^3\;\mathrm{max}(v_\sigma,\,M_N)$. Any limit derived for such a high scale thus becomes particularly sensitive to the assumption of a standard thermal history. Discussions on the effect of non-standard cosmologies and lowered reheating temperatures can be found in Ref.~\cite{Herbermann:2025uqz}.

\subsection{Model II: Enhanced Diracon model}
In the enhanced Diracon model, the interactions of charged leptons with the Diracon are not neutrino mass suppressed, unlike the model of the previous section. One can use the results in Appendix~\ref{app:coupling}  to find the general formula for these interactions, 
\begin{equation}
\mathcal{L}_{\ell\ell \D} =- \frac{i \, \D}{32\pi^2  v_\sigma} \bar{\ell} \, \left[M_\ell ~ \textup{Tr}(Y \, Y^\dagger)\, \gamma_5 +2 M_\ell\,  Y \, Y^\dagger \, P_L-2  Y \, Y^\dagger \, M_\ell \, P_R\right]\ell \, .
\end{equation}
While we will perform a general numerical scan, for illustrative purposes we again consider the degenerate BSM limit, where the matrices $M_2 = m_2 \, \id_3$ and $Y' = y' \, \id$ are degenerate. Using Eq.~\eqref{eq:Casasenhanced} under this simplifying assumption and neglecting the electron mass over the muon one we obtain

\begin{equation}
    \text{BR}(\mu \to e \, \D) = 4 \left(\frac{1}{8 \pi}\right)^5 \frac{m_\mu^3}{v^2 \Gamma_\mu} \left(\frac{m_N}{m_2}\right)^4 \frac{\left|\left(U_\ell \widehat{M_\nu}^2 U_\ell^\dagger\right)_{21}\right|^2}{v^2 v_\sigma^2}
    \label{eq:mueDenhanced}
\end{equation}
and
\begin{equation}
    S_{11} = i \frac{m_e}{16 \pi^2}\left(\frac{m_N}{m_2}\right)^2 \frac{1}{v^2 v_\sigma} \left[2 \left(U_\ell \widehat{M_\nu}^2 U_\ell^\dagger\right)_{1\,1}-\sum_i m_i^2\right] \, ,
    \label{eq:S11enhanced}
\end{equation}
which is constrained by Eq.~\eqref{eq:stellarcooling}. Note that both Eq.~\eqref{eq:mueDenhanced} and \eqref{eq:S11enhanced} depend on the same BSM factor $m_N^2 / (m_2^2 v_\sigma)$.

Comparing the most promising charged lepton decay channels, in this simplified limit we can write down
\begin{equation}
   \frac{\text{BR}(\ell_\alpha \to \ell_\beta \, \D)} {\text{BR}(\ell_\alpha \to \ell_\beta\, \gamma)} \simeq \frac{1}{16 \pi^{3} \alpha_w^{3} s_w^2}  \left(\frac{M_W^2}{v \, m_{\ell_\alpha}} \right)^2 y'^2\,  \frac{m_N^2}{v^2} \, ,
\end{equation}
which, when particularized for the case of $\mu$ decays, leads to
\begin{equation}
\label{eq:ratioenhanced}
 \frac{\text{BR}(\mu \to e \, \D)}{\text{BR}(\mu \to e\, \gamma)} \approx 2.3 \cdot\, 10^{8} \,  y'^2 \,  \left(\frac{m_N}{\text{TeV}}\right)^2 \, .
\end{equation}
In contrast to what we found in the canonical model, this ratio can now be very large, thus making the search for $\mu \to e \D$ very promising. Still, since the experimental constraints on the $\gamma$ decay are around $8$ orders of magnitude stronger than those of the $\D$, it is still possible to find testable $\mu \to e \gamma$ rates for sufficiently small Yukawas. 

Relaxing the assumption of diagonal and degenerate BSM parameters, we now proceed to perform more general numerical scans, which qualitatively still maintain the general conclusions shown above. 

\begin{figure}
    \centering
    \includegraphics[width=0.7\linewidth]{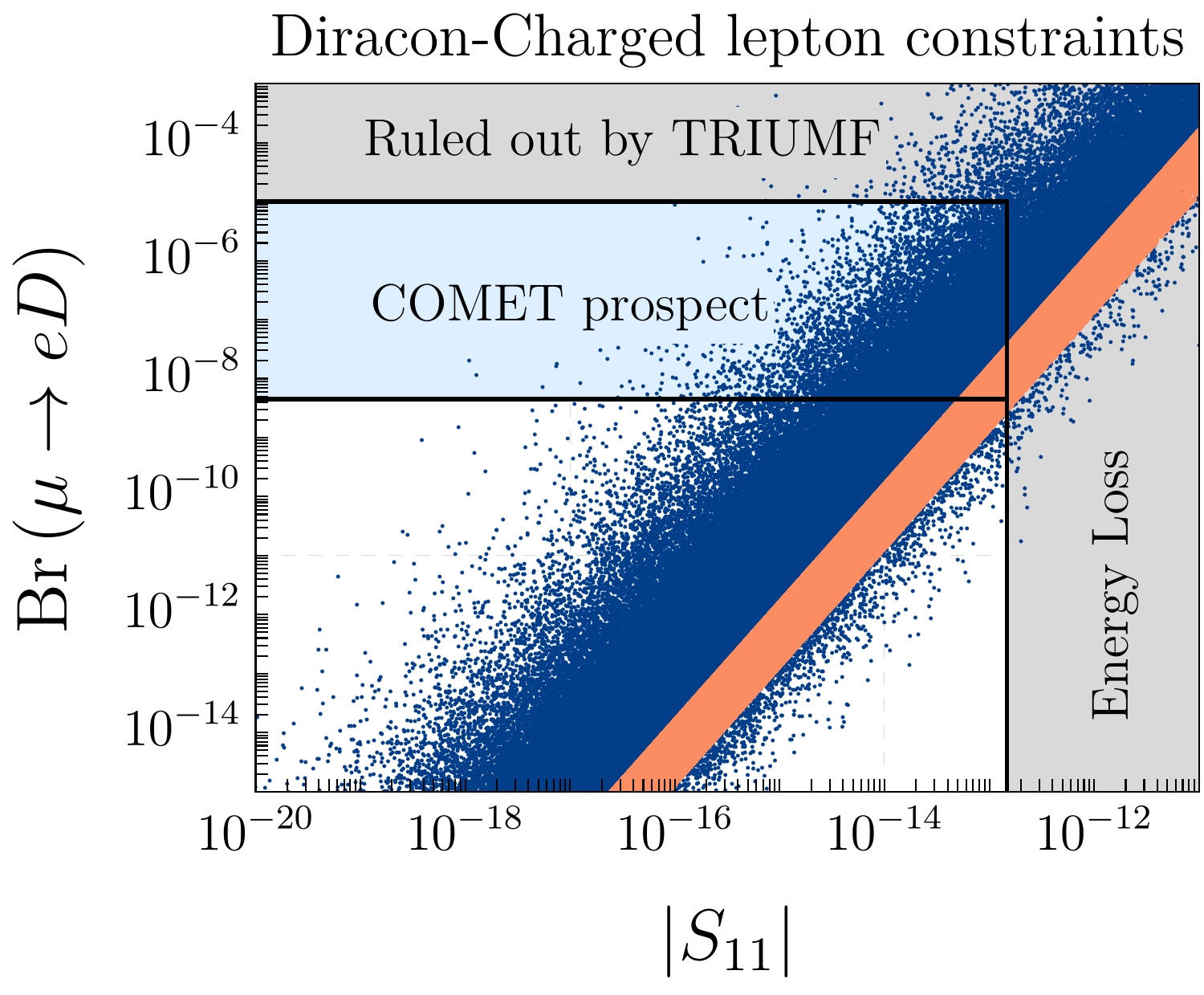}
    \caption{Enhanced Diracon model predictions for the Diracon-Charged lepton interactions, mainly constrained by muon decays, Eqs.~\eqref{eq:majoronlimit}-\eqref{eq:Comet}, and energy loss astrophysical mechanisms, Eq.~\eqref{eq:stellarcooling}. All the points shown satisfy neutrino masses and mixing from the neutrino global fit \cite{deSalas:2020pgw}, sum of neutrino masses \cite{DESI:2025zgx,ACT:2025tim}, $\text{Br}\left(\mu \to e \gamma\right)$ constraints from MEG II as well as having perturbative Yukawas and satisfying the seesaw conditions discussed in Sec.~\ref{sec:Type-I}. The orange dots follow the simplifying degenerate BSM limit while the blue ones deviate from it. The model could feature observable $\mu \to e \D$ rates while satisfying current constraints.}
    \label{fig:Diracon}
\end{figure}

In Fig.~\ref{fig:Diracon} we show the relation between the quantity $|S_{11}|$, constrained by Eq.~\eqref{eq:stellarcooling}, and the Branching ratio of the Diracon muon decay $\mu \to e \D$. As seen from Eqs.~\eqref{eq:mueDenhanced} and \eqref{eq:S11enhanced}, represented in the plot by the orange line, we expect both quantities to be highly correlated and the width related to different values of the lightest neutrino mass, $m_\text{lightest}$. While the decay $\mu \to e \D$ would be very hard to observe in COMET in the degenerate BSM limit while satisfying at the same time Eq.~\eqref{eq:stellarcooling}, deviating from this simplified assumption allows for observable muon decay rates, as seen in the blue points. All the shown points in this plot satisfy the $\mu \to e \gamma$ MEG II constraint, neutrino masses and mixing, sum of neutrino masses as well as perturbativity of the Yukawas and non-unitarity constraints.

\begin{figure}
    \centering
    \includegraphics[width=0.7\linewidth]{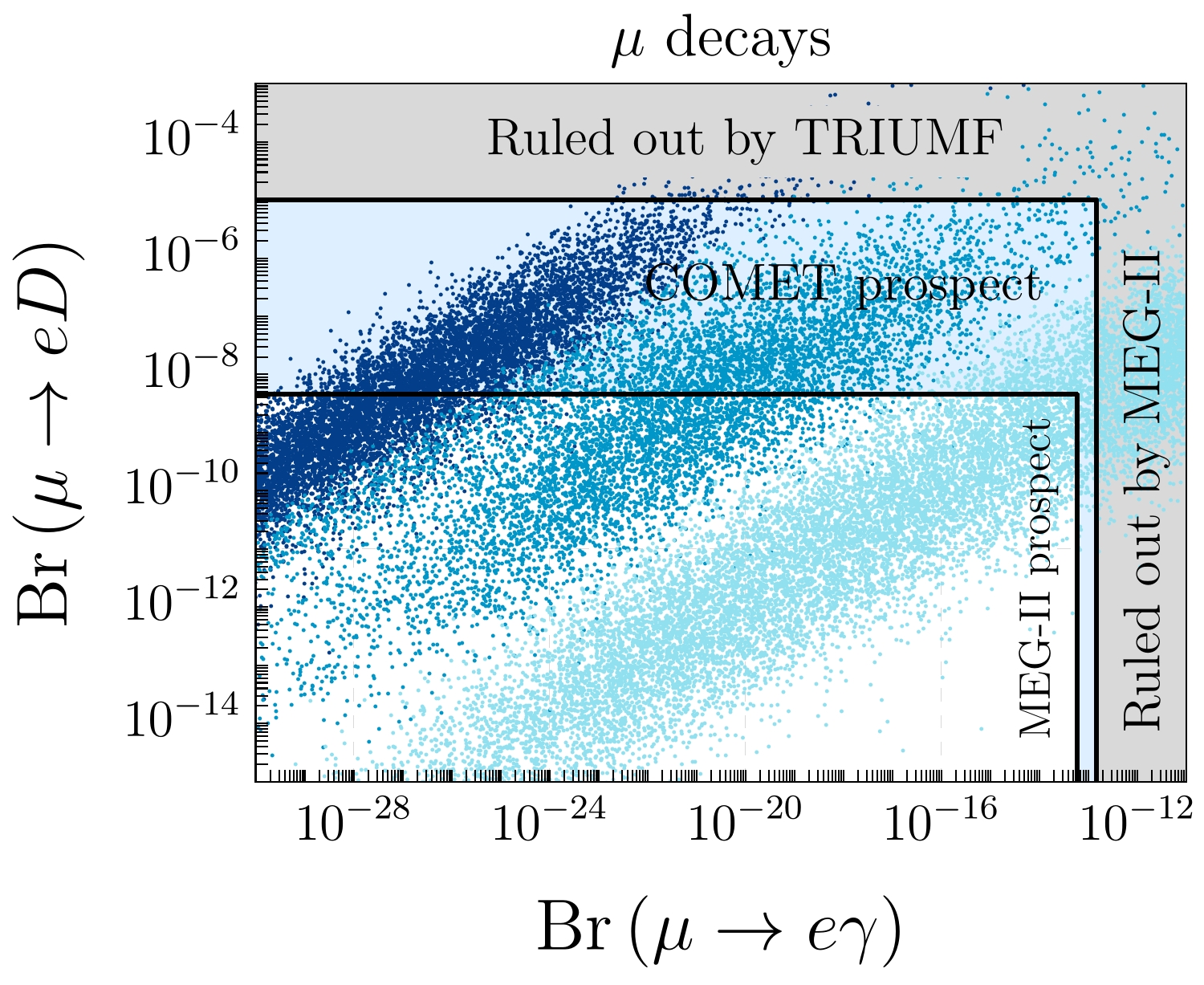}
    \caption{Comparison of the two most important BSM $\mu$ decays within the Diracon enhanced model. All the points shown satisfy neutrino masses and mixing from the neutrino global fit \cite{deSalas:2020pgw}, sum of neutrino masses \cite{DESI:2025zgx,ACT:2025tim}, as well as the constraints arising from energy loss mechanisms in astrophysical observations Eq.~\eqref{eq:stellarcooling}, perturbative Yukawas and satisfying the seesaw conditions discussed in Sec.~\ref{sec:Type-I}. The dark blue dots feature sizeable Yukawas, with $\text{Tr}(Y Y^\dagger)\, \& \,\text{Tr}(Y' Y'^\dagger) > 1$. In this regime the Diracon decay strongly dominates compared to the photon decays. Relaxing this condition and allowing for smaller Yukawas could allow observable rates of both decays at the same time, as seen in the blue dots, with $10^{-2} < \text{Tr}(Y Y^\dagger)\, \& \,\text{Tr}(Y' Y'^\dagger) < 1$, and the light blue dots, satisfying $10^{-4} < \text{Tr}(Y Y^\dagger)\, \& \,\text{Tr}(Y' Y'^\dagger) < 10^{-2}$}
    \label{fig:mudecays}
\end{figure}

On the other hand, in Fig.~\ref{fig:mudecays} we compare the rates of the two main muon decay channels, $\mu \to e \D/\gamma$, for different scales of the Yukawas $Y$ and $Y'$. As can be argued from Eqs.\eqref{eq:mueDenhanced} and \eqref{eq:ratioenhanced}, if both Yukawa matrices are of $\sim O(1)$ then we expect the Diracon decay to be observed before the $\gamma$ decay. In order to have observable $\mu \to e \gamma$ instead smaller Yukawas are needed, and we find both processes to be potentially observable in the near future if both Yukawas are of order $10^{-2}-10^{-4}$. While a more detailed analysis could be performed, it is already clear that the model has a very rich lepton flavour phenomenology.

Finally, unlike the low energy flavour-violating observables, the cosmological probes are not expected to be relevant for this model. The scalar sector is formally the same as in the canonical model. As such, the same discussion of Sec.~\ref{sec:PhenoCanonical} around Fig.~\ref{fig:scalar_limits} applies, the main difference being that in this model the VEV of $\sigma$ is responsible for the mass of the heavy neutral lepton and therefore cannot be taken to be too low. Far-future experiments like CMB-HD may constrain the $v_\sigma > v$ region of Fig.~\ref{fig:scalar_limits} in a model-independent way, see the discussion in Sec.~\ref{sec:Cosmo}.

In the neutrino sector, the $\nu_R$ production will be neutrino mass suppressed. Indeed, in the mass basis, the Yukawa term between the heavy fermions $N$ and the right-handed neutrino $\nu_R$ is of the form
\begin{equation}
    \sum_{\phi \in \{h, \mathcal{S}, \D\}}  C_\phi \widehat{ \bar{N}_L} \phi \frac{M_2}{v_\sigma} \widehat{\nu_R} \, ,
\end{equation}
where $C_\phi$ is either $\sin \alpha$ or $\cos \alpha$ times a constant of modulus $1$. The wide hat denotes the mass basis, i.e. the flavour-basis fields rotated by $U_L$ or $U_R$. Using the neutrino mass formula of Eq.~\eqref{eq:Casasenhanced} we can write
\begin{equation}
    \frac{M_2}{v_\sigma} = - Y' Y^{-1} \frac{M_\nu}{v} \, ,
\end{equation}
which is explicitly neutrino-mass suppressed. While it could be sizable for small enough $Y$, instead this would lead to a suppression of the production of $N$. Finally, the Diracon production from heavy fermion decays is not suppressed, and offers additional processes to excite the Diracon mode. However, the scalar contribution does not exceed that of a single Nambu-Goldstone boson, and for a benchmark value of $\dneff =0.02$ the forecasted cosmological limit can be summarized with an effective energy scale. To this end, we observe that production predominantly proceeds from decay of heavy $N$, and the matrix element approximately scales as $\propto (Y v/v_\sigma)^2/M_N $. From $M_N=Y^\prime v_\sigma/\sqrt{2}$, we find
\begin{equation}
    \Lambda_\mathrm{enh}^{-1} = \frac{y^2\sqrt{2} v^2}{y^\prime v_\sigma^3} \approx 10^{-15} \mathrm{\,TeV^{-1}}\,,
\end{equation}
where the numerical value corresponds to $\dneff = 0.02$ in absence of additional scalar sector contributions.
We therefore conclude that flavour observables are a more promising signature of this model over cosmological probes.







\section{Summary and Discussion}
\label{sec:summary}

We have proposed and analyzed two minimal extensions of the Standard Model in which a global, chiral and anomaly-free $U(1)_D$ symmetry is spontaneously broken, giving rise to Dirac neutrino masses and a massless Goldstone boson—the Diracon—as a physical remnant of the symmetry breaking. In both models, the smallness of neutrino masses is explained via a Dirac seesaw mechanism. The chiral nature of the $U(1)_D$ forbids the Majorana mass terms, the tree-level Dirac mass term $\bar{L} \tilde{H} \nu_R$ and, while we do not explicitly consider it, could also protect the stability of a dark matter candidate in a minimal extension of the models~\cite{CentellesChulia:2016rms, CentellesChulia:2018gwr}.

Both models have formally the same particle content and scalar potential. However, despite their similarities, the different symmetry breaking patterns lead to crucial phenomenological differences. Indeed, in the first model that we analyze, which we call the canonical model, the Diracon interactions with charged leptons are neutrino mass suppressed and thus flavour violating observables such as $\mu \to e \D$ are not expected to reach observable rates. However, it is possible to significantly produce not only the Goldstone, but also $\nu_R$ in the early universe, predominantly from decay of heavy fermions $N$. The interplay between $N$ production (driven by $Y$), and $N$ decay into $\nu_R$ (driven by $Y'$) and the neutrino mass relation $M_\nu \propto Y Y'$ leads to stringent constraints on the Yukawa sector from $N_{\text{eff}}$ and the main expected signature will be a non-zero $\Delta N_{\text{eff}}$ in the upcoming experiments. While traditional flavour observables like $\mu \to e \gamma$ are in principle possible in this framework, current and future $\Delta N_{\text{eff}}$ constraints cut much of the relevant parameter space.

In the second model, the enhanced Diracon type-I seesaw model, the situation is the opposite: the $\nu_R$ production in the early universe is $m_\nu$ suppressed. While efficient production of Diracons is possible, their abundance will remain below anticipated detection thresholds of near future experiments. Thus, the most promising signatures are the flavour observables. Naively, we expect $\mu \to e \D$ to be observed before $\mu \to e \gamma$, although the opposite situation is possible by considering small Yukawas.

The two models are well motivated and minimalistic extensions of the SM which  explicitly show the complementarity between cosmological and low energy probes in the search for BSM physics. While we do not discuss the following topics in this manuscript, open directions for future works include performing a detailed collider phenomenology analysis, non-standard neutrino propagation due to neutrino mixing with HNL, the possible role of the Diracon as a dark matter component if the $U(1)_D$ symmetry is explicitly broken, and analyzing leptogenesis within the models, see e.g. \cite{Dick:1999je,Murayama:2002je,Cerdeno:2006ha,Heeck:2023soj}.

\section*{Acknowledgements}

Work supported by the Spanish grants PID2023-147306NB-I00, CNS2024-154524 and CEX2023-001292-S (MICIU/AEI/10.13039/501100011033), as well as CIPROM/2021/054 (Generalitat Valenciana). The work of AHB is supported by the grant
No. CIACIF/2021/100 (also funded by Generalitat Valenciana). TH acknowledges support from the IMPRS-PTFS. SCCh acknowledges support from the MPIK, where this work started, and the Spanish grant CIPROM/2021/054 (Generalitat Valenciana). TH would like to thank Paul Frederik Depta for valuable comments on the numerical implementation.
\appendix

\section{The Diracon coupling to charged leptons}
\label{app:coupling}

Following the notation in~\cite{Herrero-Brocal:2023czw}, we generally express the interaction of the Diracon with a pair of charged leptons as~\cite{Escribano:2020wua}
\begin{equation} \label{eq:llD}
\mathcal{L}_{\ell \ell \D} = \D \, \bar{\ell}_\beta \left[ S^{\beta \alpha} \, P_L + \left( S^{\alpha \beta} \right)^* \, P_R \right] \ell_{\alpha} \, .
\end{equation}
Here $P_{L,R} = \frac{1}{2} \left( 1 \mp \gamma_5 \right)$ are the usual chiral projectors, $\ell_{\alpha,\beta}$ are the charged leptons, with $\alpha,\beta$ two generation indices, and $S$ is expressed in terms of the contributions from the $Z$ and $W$ bosons ($\Gamma_Z$ and $L_W$, respectively),
\begin{equation} \label{eq:Scoup}
S^{\beta \alpha} = \frac{1}{8 \pi^2}\left( \delta^{\beta \alpha} \, \Gamma_Z^\alpha + L_W^{\beta \alpha} \right) \, .
\end{equation}
Modifying the general result in \cite{Herrero-Brocal:2023czw} to introduce the Dirac nature of neutrinos, we can derive the dominant terms for $\Gamma_Z^\alpha$ and $L_W^{\beta \alpha}$. Before presenting the result let us set our notation, which differs from the one used in the original reference. We use $M_\ell = \textup{diag}\left( m_{\ell_e}, \, m_{\ell_\mu}, \, m_{\ell_\tau} \right)$ for the charged lepton mass matrix. We also need to define the coupling between the Diracon and the neutrinos in the flavour basis, $\Y$, defined by the following Lagrangian term, 
\begin{equation}
    \L \subset  \D \, \begin{pmatrix}
      \bar{\nu}_L &   \bar{N}_L
    \end{pmatrix} \Y \begin{pmatrix}
        \nu_R \\
        N_R
    \end{pmatrix}  + \hc \, ,
\end{equation}
or, equivalently, 
\begin{equation}
    \L \subset  \D \, \bar{n} \left( U_L^\dagger \, \Y \, U_R \, P_R + U_R^\dagger \, \Y^\dagger \, U_L \, P_L \right) n \, .
\end{equation}
Finally, just as in the original reference, we use $\sum_{j \sim l}$ and $  \sum_{j \sim h} $ to refer to the sum over light and heavy neutrinos, respectively.\\

With this notation at hand, the dominant terms for $\Gamma_Z^\alpha$ and $L_W^{\beta \alpha}$ are given by
\begin{align}
    \Gamma_Z^\alpha & \simeq -i \frac{m_{\ell_\alpha}}{6 v^2 } \IM \left[ \sum_{s=1}^3 \left( \Gamma_{ss} -2  \Delta_{ss} \right) \right] \, , \\
     L_W^{\beta \alpha} &\simeq \frac{ m_{\ell_\beta}}{12 \, v^2} \left[ \left( \Gamma_{\alpha \beta }^*+ 8 \, \Gamma_{\beta \alpha} \right) - \left( 2 \, \Delta_{\alpha \beta}^* +7 \, \Delta_{\beta \alpha} \right) \right] \, ,
\end{align}
where $\Gamma$ and $\Delta$ are defined as follows:
\begin{align}
   \Gamma_{\beta \alpha} &= \sum_{k,r} \Y_{kr} \M^\dagger_{r \alpha } \sum_{j \sim l} \left(U_L \right)_{\beta j} \left( U_L^\dagger \right)_{jk}  \, ,\\
   \Delta_{\beta \alpha} &= \sum_{k,r} \Y_{kr} \left( \M \M^\dagger \right)_{\beta k} \sum_{j \sim h} \left(U_R\right)_{r j} m_j^{-1} \left( U_L^\dagger \right)_{ j \alpha } \, .
\end{align}

\section{Scalar interactions}
\label{app:scalar}
In general, for a theory with three scalars, $\phi$, $\varphi$ and $\sigma$ amplitudes can be written as
\begin{align}
\mathcal{A}(\phi \phi \to \varphi \varphi) =&  \lambda_{\phi^2 \varphi^2} + \lambda_{\phi^3}\lambda_{\phi \varphi^2} P(s, m_\phi)+ \lambda_{\varphi^3}\lambda_{\phi^2 \varphi} P(s, m_\varphi) + \lambda_{\phi^2 \sigma} \lambda_{\sigma \varphi^2} P(s, m_\sigma)\nonumber \\
&+ \lambda_{\phi \varphi^2}^2 \left( P(t, m_\varphi)+ P(u, m_\varphi) \right) + \lambda_{\phi^2 \varphi}^2 \left( P(t, m_\phi) + P(u, m_\phi) \right) \nonumber \\
&+ \lambda_{\phi \sigma \varphi}^2 \left( P(t, m_\sigma) + P(u, m_\sigma) \right) \, , \\
 \mathcal{A}(\phi \to \varphi \varphi)  =& \lambda_{\phi \varphi^2} \, ,
\end{align}
with $s=\left( p_1 + p_2 \right)^2 $, $t=\left( p_1 - p_3 \right)^2$ and $u=\left( p_1 - p_4 \right)^2$ the Mandelstam variables and $P(p,m_i)$ the propagator, 
\begin{equation}
P(p,m_i) = \frac{1}{p-m_i^2 + i \, m_i \, \Gamma_i} \, .
\end{equation}
$\Gamma_i$ is the total decay width. In the completely broken phase, we make the identification $h = \phi $ , $\varphi = J \, , \mathcal{S} $, $\sigma = \mathcal{S}\, , J $;  $h = \cos \alpha \,  H +\sin \alpha \,  \sigma$,  $\mathcal{S} = -\sin \alpha \,  H +\cos \alpha \,  \sigma$. If one or both symmetries are in their unbroken phase, the field assignment needs to be adapted accordingly, and a similar expansion of the scalar portal gives rise to the appropriate coupling expressions.

The partial decay width of one scalar to two is given by
\begin{equation}
    \Gamma(\phi \to \varphi \varphi) = \frac{\kappa_{\phi \varphi^2}^2}{8 \pi m_\phi} \sqrt{1-\frac{4 m_\varphi^2}{m_\phi^2}} \, , \label{eq:scadecay}
\end{equation}
while the partial decay width of one scalar to two fermions (in the absence of a Dirac matrices structure and with Yukawa coupling $Y$) is given by
\begin{equation}
\Gamma(\phi \to \bar{\psi} \psi) =  \frac{Y^2 m_\phi}{8 \pi} \left(1- 4 \frac{m_\psi^2}{m_\phi^2}\right)^{3/2} \, .
\end{equation}

We now express all parameters in the potential in terms of $\alpha$, $v$, $v_\sigma$, and the scalar masses, which allows us to read off the relevant couplings. We perform a leading order expansion in $\frac{v}{v_\sigma} \sin \alpha  \ll 1$, which we can safely assume due to Higgs invisible decay limits, as well as $\sin \alpha \ll 1$. 

At first order, the cubic couplings of mass dimension are given by ($\tan \beta = v/v_\sigma$)
\begin{align}
    \kappa_{h^3} &= \frac{\sqrt{G_F} m_h^2}{2^{3/4}} \, , \\
    \kappa_{h^2S} &= \frac{\sin \alpha  \sqrt{G_F}}{2^{3/4}} \left(2 m_h^2+m_\mathcal{S}^2\right) \, , \\
    \kappa_{h S^2} &= -\frac{\sin \alpha  \sqrt{G_F} \tan \beta}{2^{3/4}}\left(m_h^2 + 2 m_\mathcal{S}^2\right) \, , \\
    \kappa_{S^3} &= \frac{\sqrt{G_F} m_\mathcal{S}^2 \tan \beta}{2^{3/4}} \, , \\
    \kappa_{hDD} &= -\frac{\sin \alpha  \sqrt{G_F} m_h^2 \tan \beta}{2^{3/4}} \, , \\
    \kappa_{SDD} &= \frac{\sqrt{G_F} m_\mathcal{S}^2 \tan \beta}{2^{3/4}} \,.
\end{align}
Similarly, sorting terms in the scalar potential yields for the quartic couplings
\begin{align}
    \lambda_{h^4} &= \frac{G_F m_h^2}{4 \sqrt{2}} \, , \\
    \lambda_{h^3S} &= \frac{\sin \alpha  G_F m_h^2}{\sqrt{2}} \, , \\
    \lambda_{h^2S^2} &= -\frac{\sin \alpha  G_F \tan \beta \left(m_h^2-m_\mathcal{S}^2\right)}{2 \sqrt{2}} \, , \\
    \lambda_{hS^3} &= -\frac{\sin \alpha  G_F m_\mathcal{S}^2 \tan^2\beta }{\sqrt{2}} \, , \\
    \lambda_{S^4} &= \frac{G_F m_\mathcal{S}^2 \tan^2\beta }{4 \sqrt{2}} \, , \\
    \lambda_{h^2D^2} &= -\frac{\sin \alpha  G_F \tan \beta \left(m_h^2-m_\mathcal{S}^2\right)}{2 \sqrt{2}} \, , \\
    \lambda_{hSD^2} &= -\frac{\sin \alpha  G_F m_\mathcal{S}^2 \tan^2\beta }{\sqrt{2}} \, , \\
    \lambda_{S^2D^2} &= \frac{G_F m_\mathcal{S}^2 \tan^2\beta }{2 \sqrt{2}} \, , \\
    \lambda_{D^4} &= \frac{G_F m_\mathcal{S}^2 \tan^2\beta }{4 \sqrt{2}} \, .
\end{align}

\section{Yukawa interactions}
\label{app:Yukawa}
In the unbroken phase, Yukawa couplings can directly be read off the Lagrangian. In the broken phase after SSB, both models can be handled simultaneously. For the sake of generality we write the scalar-fermion-fermion interaction terms as 
\begin{equation}
    \L_{\textup{int}} = \begin{pmatrix}
        \bar{L} & \bar{N}_L
    \end{pmatrix}\cdot \begin{pmatrix}
        0 & \frac{1}{\sqrt{2}} \, Y S_h\\
        \frac{1}{\sqrt{2}} Y_2 \left(S_\sigma + i \D\right) & \frac{1}{\sqrt{2}} Y_M \left(S_\sigma + i \D\right)
    \end{pmatrix} \cdot \begin{pmatrix}
        \nu_R \\ N_R
    \end{pmatrix} +\hc \, .
\end{equation}
Here $S_h = h \cos \alpha + \mathcal{S} \sin\alpha$, $S_\sigma = \mathcal{S} \cos \alpha - h \sin\alpha$. In the canonical model we can identify $Y_2=Y'$ and $Y_M=0$, and in the enhanced Diracon model $Y_2=0$ and $Y_M=Y'$. We now rotate to the mass basis, obtaining
\begin{equation}
    \L_{\textup{int}} = \begin{pmatrix}
        \overline{\widehat{\nu}}_L & \overline{\widehat{N}}_L
    \end{pmatrix}\cdot U_L^\dagger \cdot \begin{pmatrix}
        0 & \frac{1}{\sqrt{2}} \, Y S_h\\
        \frac{1}{\sqrt{2}} Y_2 \left(S_\sigma + i \D\right) & \frac{1}{\sqrt{2}} Y_M \left(S_\sigma + i \D\right)
    \end{pmatrix} \cdot U_R \cdot \begin{pmatrix}
        \widehat{\nu}_R \\ \widehat{N}_R
    \end{pmatrix} +\hc \, ,
\end{equation}
where the wide hat represents the field in the mass basis. Taking Eq.~\eqref{eq:Ufactor}, substituting $P_L=M_1 M_N^{-1}$, $P_R^\dagger = M_N^{-1}M_2$ and $U_{hL} = U_{hR} = I$ and neglecting factors of the order of the neutrino mass or lower we find for the scalar-heavy-light interactions
\begin{equation}
    \mathcal{L}_{int} \supset \overline{\widehat{\nu}}_L \, U_{lL}^\dagger \left(S_h \frac{Y}{\sqrt{2}} - \frac{S_\sigma}{\sqrt{2}} M_1 M_N^{-1} Y_M\right) \, \widehat{N}_R + \overline{\widehat{N}}_L\,\frac{S_\sigma}{\sqrt{2}}\left(Y_2 - M_2 M_N^{-1} Y_M\right) \, U_{lR}  \, \widehat{\nu}_R \, .\label{eq:interactions}
\end{equation}

In the canonical version we make the replacements $M_1 = \frac{1}{\sqrt{2}} Y v$, $M_2 = \frac{1}{\sqrt{2}} Y' v_\sigma$, $Y_2= Y'$ and $Y_M=0$. In the enhanced Diracon model, instead, we have $M_1 = \frac{1}{\sqrt{2}} Y v$, $M_N = \frac{1}{\sqrt{2}} Y' v_\sigma$, $Y_M= Y'$ and $Y_2=0$.
Furthermore, we can use Eq.~\eqref{eq:Casasenhanced} to rewrite $\mu/v_\sigma \, U_{lR}$ as
\begin{equation}
\frac{\mu \, U_{\ell R}}{v_\sigma} =\frac{1}{v_H} \, Y' \, Y^{-1} \, U_\ell \, \widehat{M}_\nu \, ,
\end{equation}
to explicitly show that this quantity is neutrino-mass suppressed. A summary of the resulting vertices is presented in Tab.~\ref{tab:Yukawa_vertices}.

\begin{table}[]
    \centering
    \begin{tabular}{|c|c|c|}
    \hline
        Vertex & Canonical & Enhanced Diracon \\
        \hline
        $N_R \, \nu_L \, h$ & $U_{lL}^\dagger Y \cos \alpha / \sqrt{2}$ &  $U_\ell^\dagger \, Y (\frac{v}{v_\sigma} \sin \alpha+ \cos \alpha)/ \sqrt{2}$\\
        $N_R \,\nu_L \,S$ & $U_{lL}^\dagger Y \sin \alpha/ \sqrt{2}$ & $U_\ell^\dagger \, Y (\sin \alpha - \frac{v}{v_\sigma} \cos\alpha)/ \sqrt{2} $ \\
        $N_R \,\nu_L \,\D$ & $0$ & $-i \frac{v}{v_\sigma} U_\ell^\dagger \, Y / \sqrt{2}$ \\
        $N_L \,\nu_R\, h$ & $-Y'U_{lR} \sin \alpha/ \sqrt{2}$ & $\frac{\mu}{v_\sigma}  U_{lR}\sin \alpha$\\
        $N_L \,\nu_R \,S$ & $Y'U_{lR} \cos \alpha/ \sqrt{2}$ & $-\frac{\mu}{v_\sigma} U_{lR} \cos \alpha$ \\
        $N_L \,\nu_R \,\D$ & $i Y' U_{lR}/ \sqrt{2}$ & $-i \frac{\mu}{v_\sigma}  U_{lR}$ \\
        \hline
    \end{tabular}
    \caption{Yukawa couplings in the broken phase after rotation to the mass basis.
    \label{tab:Yukawa_vertices}}
\end{table}

\bibliographystyle{utphys2}
\providecommand{\href}[2]{#2}\begingroup\raggedright\endgroup

\end{document}